\documentclass{jfm}
\usepackage{graphicx}
\usepackage{epstopdf, epsfig}
\usepackage{amsmath}
\usepackage{amssymb}
\usepackage{mathtools}
\usepackage{xcolor}

\shorttitle{Shear instability of an axisymmetric air-water coaxial jet}
\shortauthor{J.-P. Matas, A. Delon and A. Cartellier}

 \title{Shear instability of an axisymmetric air-water coaxial jet}
 \author{Jean-Philippe Matas\aff{1}
  \corresp{\email{jean-philippe.matas@univ-lyon1.fr}}, Antoine Delon\aff{2} \and Alain Cartellier\aff{2}}

\affiliation{\aff{1}Laboratoire de M\'ecanique des Fluides et d'Acoustique, Ecole Centrale de Lyon, CNRS, Universit\'e Claude Bernard Lyon 1, INSA Lyon, F-69134 Ecully, France
\aff{2}Univ. Grenoble Alpes, CNRS, Grenoble INP\thanks{Institute of Engineering Univ. Grenoble Alpes}, LEGI, F-38000 Grenoble, France}

\begin{document}

\maketitle

\begin{abstract}
We study the destabilization of a round liquid jet by a fast annular gas stream. We measure the frequency of the shear instability waves for several geometries and air/water velocities. We then carry out a linear stability analysis, and show that there are three competing mechanisms for the destabilization: a convective instability, an absolute instability driven by surface tension, and an absolute instability driven by confinement. We compare the predictions of this analysis with experimental results, and propose scaling laws for wave frequency in each regime. We finally introduce criteria to predict the boundaries between these three regimes.
\end{abstract}


\section{Introduction}
Liquid atomization is the process by which a liquid stream is destabilized and converted into a spray. This fragmentation can be efficiently carried out with the help of a gas stream: airblast atomization is the process by which the liquid bulk is fragmented intro droplets by a fast
parallel gas jet \citep{Lefebvre}. Drop fragmentation in this configuration is the result of several instabilities affecting the liquid stream \citep{Lasheras, Marmottant, Hong2002, Ling2017}: a shear instability leads to the formation of two dimensional waves at the surface of the liquid jet ; the crest of these waves destabilizes and is stretched into ligaments or thin sheets by the fast gas stream ; ligaments/sheets eventually fragment under the action of capillary forces and/or gas Reynolds stresses. The aim of the present paper is to describe the initial shear instability, and clarify its nature in the range of parameters relevant to the atomization of an initially round axisymmetric liquid jet by a coaxial annular gas jet. 

It has been shown that the order of magnitude and variations of the experimental frequency and velocity of these waves could be captured by a simple inviscid linear stability analysis \cite{Raynal_turb, Marmottant}, provided the velocity deficit introduced by the splitter plate is taken into account in the base flow \citep{Matas}. It was however observed that predicted frequencies were smaller than experimental ones by a factor three in the round coaxial jet case. It was  demonstrated a few years later that for certain conditions experimental frequencies could be predicted  accurately by a (much more complex) spatio-temporal viscous stability analysis \citep{Otto, Fuster}. It was shown that depending on conditions, including the magnitude of the velocity deficit in the base flow, the mechanism could be either convective or absolute. The instability mechanism at work in these cases was argued to be a viscous one, the reason being that the shear instability branch involved in all these cases corresponds to a mode driven by viscous stresses at the interface \citep{Yih,Hooper}. The absolute instability cases correspond to a pinching of this branch with a surface tension controlled branch. It was simultaneously observed in air-water experiments and numerical simulations at a reduced density ratio that the boundary between these convective and absolute regimes was controlled by the dynamic pressure ratio $M=\rho_G U_G^2/\rho_LU_L^2$, where $\rho_G$ and $\rho_L$ are respectively the gas and liquid densities, and $U_G$ and $U_L$ the gas and liquid mean velocities \citep{Fuster}.

More recently, it was pointed out by \citet{Matas_confi} that inclusion of confinement in this problem led to the occurrence of a different type of absolute instability, following the mechanism introduced by \citet{Healey} and \citet{Juniper2007}: the shear instability branch resonates with a confinement branch. The pinch point characteristic of this absolute instability is located at much lower wavenumbers than for the surface tension mechanism. An energy budget shows that in this case the kinetic energy of the perturbation is mostly fed by gas Reynolds stresses, and that the mechanism can therefore be considered as globally inviscid. This mechanism, which explains the relevance of the initial simplified inviscid analysis, seems to take over the viscous one for larger gas velocities, but the exact conditions for which it occurs are unknown. While a cartography of the different dominating destabilizing mechanisms has been established for the case of two-layer channel flow \citep{ONaraigh}, there is no such comprehensive work in the case of coaxial jets.

The first objective of the present paper is to determine if the confinement mechanism, which has been validated for planar geometry mixing layers (theory + experiments), can be transposed to an axisymmetric geometry, where confinement issues are a priori different since there is no solid wall. A second objective is to clarify the issue of the boundaries between the three observed regimes: convective, absolute due to confinement or absolute due to surface tension. Finally, the third objective is to explicit the  scaling laws for the most unstable frequency in each of these three regimes.

We will address these questions in the following manner:
\begin{itemize}
\item [-] We will present in the second section our experimental set-up and measurement techniques. We will then explain how we carry out the linear stability analysis of this problem.
\item [-] We will introduce in the third section our new experimental data for the coaxial air/water jet configuration in a wider range of experimental parameters, including new geometries. We will then compare this data to the predictions of the viscous stability analysis, and show that the same three regimes identified in the planar geometry are recovered.
\item [-] We will discuss these results in the fourth section, namely identify the boundaries between the three regimes and propose scaling laws for each of them.
\end{itemize}
 
\section{Experimental set-up and methods}

\subsection{Experimental set-up}
\begin{figure}
\includegraphics[width=13cm]{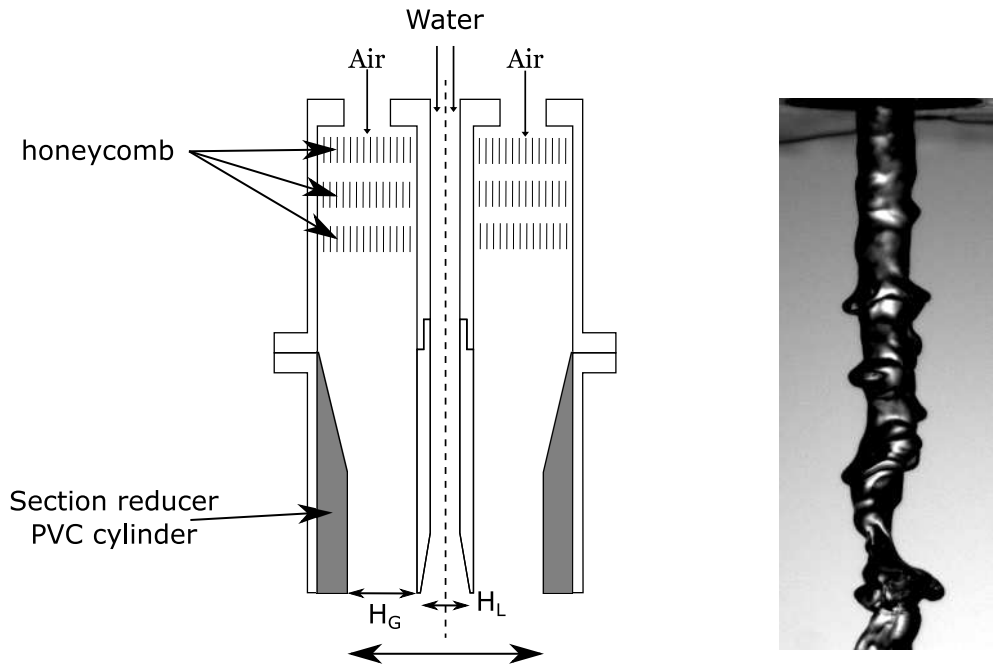}
\caption{a) Sketch of the coaxial injector. b) Typical image of the liquid jet for experimental conditions $H_L$ = 5 mm, $H_G$ = 5 mm, $U_L$ = 1.4 m/s and $U_G$ = 31 m/s. c) Example of large scale displacement of the liquid jet center for the same injector, $U_L=0.28$~m/s and $U_G=19.5$~m/s.}
\label{fig:injector}
\end{figure}

We use the axysimmetric injector of figure \ref{fig:injector}a), in which we can modify both the liquid diameter $H_L$ and the gas stream thickness $H_G$. The liquid diameter $H_L=2R$ is modified by changing the bottom part of the liquid tube. The gas thickness $H_G$ is modified by inserting PVC cylinders of varying thickness. The thickness of the inner cylinder on the exit section (lip thickness) is 200 $\mu$m for all geometries.  The injector is fed with the compressed air of the laboratory (Atlas Copco GA 11) and flow rate is controlled with a regulated pressure valve (ASCO Numatics). Water is provided by an overflowing tank in order to ensure a steady liquid flow rate.

Maximal gas velocity $U_{G0}$ is measured with a Pitot tube and a TSI manometer, with an uncertainty of 7\%. The gas flow rate and mean gas velocity $U_G$ are simultaneously measured with a mass flow meter Brooks SLA 5860. The measurement uncertainty of this device is about 3\%. Liquid mean velocity $U_L$ is measured with a gear flow meter Oval, with a maximum uncertainty of 5\%.

The gas velocity profile is measured with hot wire anemometry. These measurements are carried out without any liquid flow. We deduce from the radial velocity profiles the thickness of the vorticity layer, defined as:
\begin{equation}
\delta_G = \frac{U_{G0}}{\frac{dU_G}{dr}|_{max}}
\end{equation}
These gas velocity profiles are measured at a downstream vertical distance of 0.1 to 0.3 mm from the lip of the liquid injector. Liquid velocity profiles are measured indirectly with hot wire anemometry by feeding the central tube with air flow and using Reynolds number similarity to infer the liquid vorticity thickness $\delta_L(Re)$ for each geometry.

In order to measure the shear instability frequency, we extract the edges of the liquid jet from single-view backlit shadowgraphy and high-speed imaging. The liquid jet is placed between a high speed camera Phantom Miro 310 and a spotlight with a diffuser screen. Typical images are shown in figure \ref{fig:injector}b). The full resolution of the camera is 1280 by 800 pixels at a frequency at least equal to 1 kHz. To ensure statistical convergence, the pictures are acquired over a time equal to at least 200 periods of instability (this is verified a posteriori).

The shear instability frequency is extracted with the following image processing routine: a background subtraction is applied on pictures,  the images are thresholded based on Otsu's method, a segmentation computation is carried out in order to extract areas of the liquid jet connected to the injector exit; the radius of the jet at a given height is  extracted as the horizontal distance  between both edges of the thresholded structure ;  we  compute a fast Fourier transform of the temporal jet radius signal using Welch method, with the signal split into five segments. The most energetic frequency of the spectrum is retained as the frequency of the shear instability. This frequency is independent of the threshold introduced in the processing. 
The spectral resolution resulting from the sampling frequency and signal length is below 5 Hz for most of our measurements, and reaches 9 Hz for the largest gas velocities/frequencies investigated. The resulting relative uncertainty on the most unstable frequency is below 3\% for all conditions, except for the $H_L = 20$~mm $H_G=24$~mm geometry for which it is below 8\%.

\begin{figure}
\centering
\includegraphics[width=0.3\textwidth]{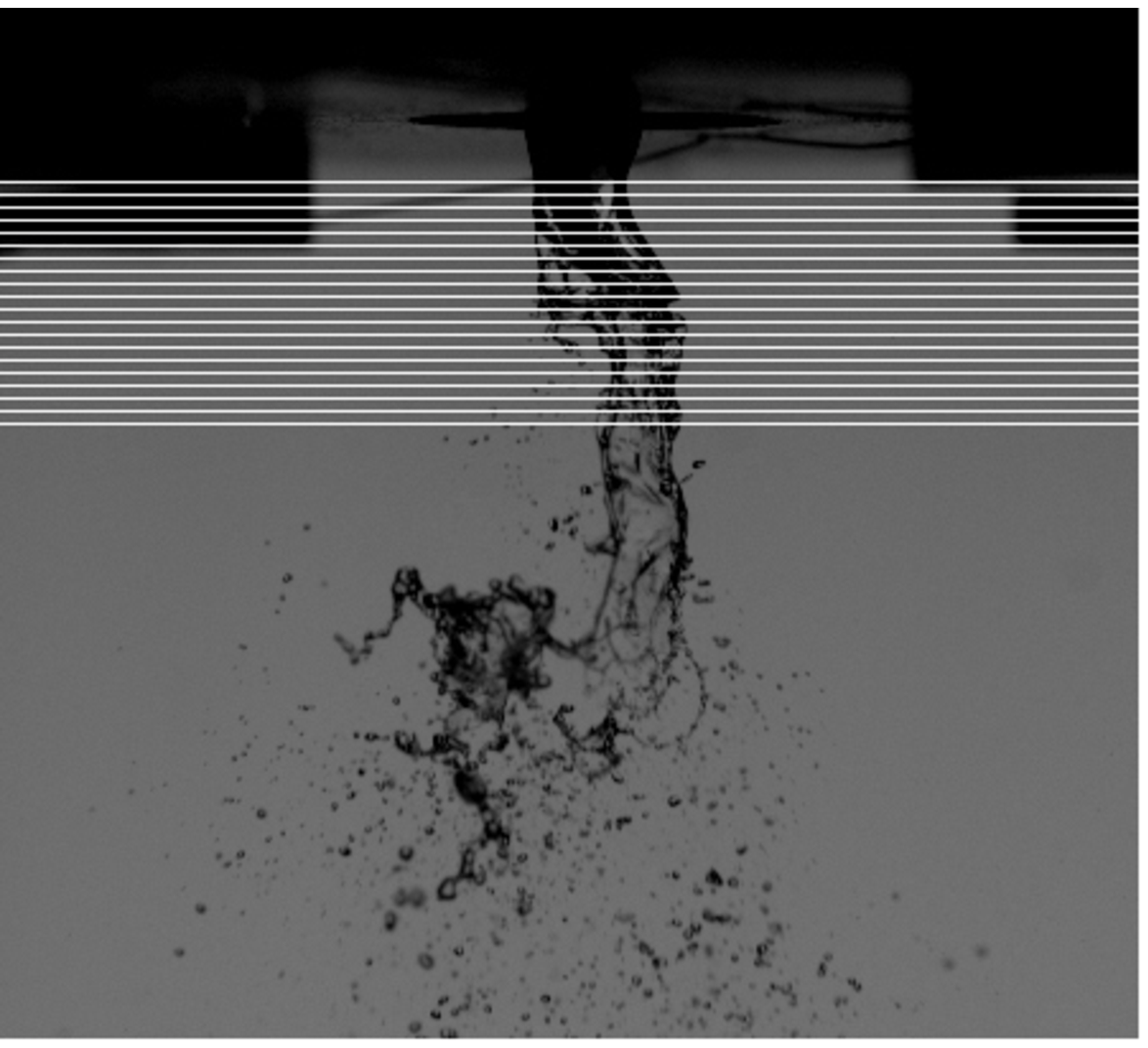}
\hspace{2cm}
\includegraphics[width=0.45\textwidth]{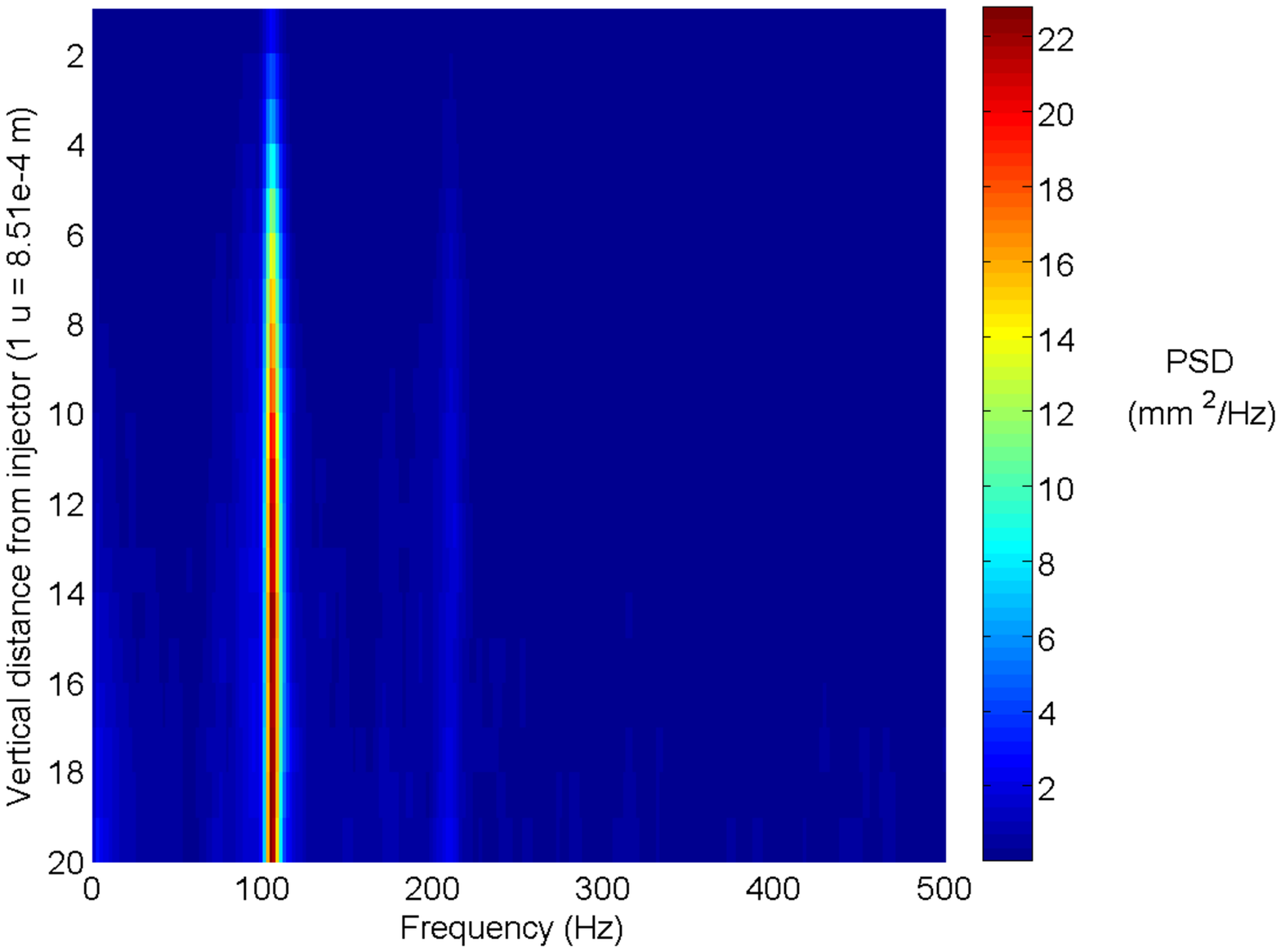}
\caption{Left: Locus of measurement of liquid jet radius spectrum for the $H_G=5$~mm $H_L=5$~mm injector, $U_L=0.28$~m/s and $U_G=15.2$~m/s; Right: Corresponding power spectrum density of jet radius $R(t)$ as a function of frequency and vertical distance $z$ for the same injector and conditions.}
\label{fig:frequ_KH_Hg5Hl5}
\end{figure}

For certain conditions, a distinct large scale instability can occur, characterized by a lateral displacement of the center of the liquid jet larger than the radius. An example of this kind of liquid structure is given in figure \ref{fig:injector}c). 
Our processing for the shear instability frequency is not affected by this lateral displacement of the liquid jet center provided the computation of the projected radius of the jet is carried out at downstream distances such that the lateral displacement is smaller than a liquid radius. This maximum downstream distance can be statistically inferred from the extracted edges of the liquid jet.
Figure \ref{fig:frequ_KH_Hg5Hl5} shows an example of spectra computation for increasing downstream distances: the maximum frequency becomes sharper as one goes downstream, and its value remains constant with downstream distance.

\subsection{Stability analysis}

We assume a base flow of the form $\mathbf{U}=(0,0,U(r))$ in a cylindrical frame ($\mathbf{e_r}$,$\mathbf{e_\theta}$,$\mathbf{e_z}$) where axis $z$ is taken along the jet axis and $r$ is the distance to this axis. The base flow is of the form suggested by \citet{Otto},  a sum of error functions accounting for the wake behind the separator plate:
$$\left \{\begin{array}{lll}
U(r) =U_{L0} \,\mathrm{erf}\left(\frac{R-r}{\delta_L}\right)
+U_i\left[1+\mathrm{erf}\left(\frac{r-R}{\delta_d\delta_L}\right)\right]
 &\mathrm{for} \; 0<r<R
\\
U(r) =\left[U_{G} \,\mathrm{erf}\left(\frac{r-R}{\delta_G}\right)+U_i\left[1-\mathrm{erf}
\left(\frac{r-R}{\delta_d\delta_L}\right)\right]\right]\left(\frac{1+\mathrm{erf}\left(\frac{H_G-r+R}{\delta_G}\right)}{2}\right)
\; &\mathrm{for} \;  R<r<L_G
\end{array}
\right.$$
where $U_{L0}$ is the liquid velocity far from the interface, $\delta_L$ the liquid vorticity thickness and $L_G$ is the radial distance at which a boundary condition with a solid wall is enforced. For the configurations explored below, we have taken $L_G>10 (R+H_G)$. The wake behind the splitter plate is modeled by the contribution proportional to the interfacial velocity $U_i$. This contribution is a vorticity layer of thickness $\delta_d\delta_L$ : $\delta _d=1$ corresponds to the absence of a velocity deficit, while $\delta_d\ll 1$ corresponds to a near zero velocity at the interface \citep{Otto}. Finally, the magnitude of the interfacial velocity $U_i$ is set by the continuity of tangential stresses: 
\begin{equation}
U_i=  \frac{U_G\mu_G/\delta_G +U_L\mu_L/\delta_L}{\mu_G+\mu_L}\delta_d\delta_L
\label{eq:Ui}
\end{equation}
where $\mu_G$ and $\mu_L$ are respectively the gas and liquid dynamic viscosities.

The general method is similar to that used in \citet{Matas_confi}, but with the cylindrical coordinate equations introduced in \citet{Matas_Hong}: we seek to determine the stability of a small velocity perturbation $\mathbf{u}(r,\theta,z,t)$ superimposed on the above velocity profile. After linearization, we expand the perturbation on Fourier modes $\tilde{\mathbf{u}}(r,n,k,\omega)e^{i(n\theta+kz-\omega t)}$. We limit ourselves to the search for axisymmetric perturbations, hence we only retain the first term in the Fourier series for $\theta$ ($n=0$). We can then introduce the stream function $\phi$ from which the respectively axial and radial velocity components $\tilde{u}$ and $\tilde{v}$ can be recovered:
\begin{align}
\tilde{u}&=\frac{1}{r}\frac{d\phi}{dr}& \tilde{v}&=-\frac{ik}{r}\phi
\end{align}
The equation for $\phi(r,k,\omega)$ is then a circular Orr-Sommerfeld equation:
\begin{align}
(Uk-\omega)&\left(\phi ''-\frac{\phi'}{r}-k^2\phi\right)+\phi k\left(\frac{U'}{r}-U''\right)
\notag
\\
=&-i\nu_{G/L}\left[\phi''''-\frac{2}{r}\phi'''+\frac{3}{r^2}\phi''-\frac{3}{r^3}\phi'-2k^2\left\{\phi''-\frac{\phi'}{r}\right\}+k^4\phi\right]
\end{align}
where $\nu_{G/L}$ is the kinematic viscosity of the gas/liquid phase. Derivatives of $\phi$ with respect to $r$ are noted with primes. We enforce boundary conditions at the outer wall: $\phi(L_G)=0$ and $\phi'(L_G)=0$, as well as on the axis of symmetry of the system $\phi(0)=0$ and $\phi'(0)=0$. Two solutions are then integrated in the gas phase from $r=L_G$ to $r=R$, and two solutions in the liquid phase from $r=0$ to $r=R$. Continuity of normal and tangential velocity and stress at the interface close the system: 
\begin{align}
\label{v_cont}
\phi_L=\phi_G
\\
\phi_G'-\phi_L'=\frac{k\phi_G}{kU_i-\omega}\left(U_G'(R)-U_L'(R)\right)
\\
\mu_G\left(\phi_G'''-\frac{\phi_G''}{R}\right)-\phi_G'\left[i\rho_G\left(kU_i-\omega\right)-\frac{\mu_G}{R^2}+3\mu_Gk^2\right]
\notag
\\
+\phi_G\left(ik\rho_G U_G'(R)+2\mu_G\frac{k^2}{R}\right)+i\sigma \frac{k^2}{R^2}\frac{1}{kU_i-\omega}\left(1-k^2R^2\right)   = \mu_L\left(\phi_L'''-\frac{\phi_L''}{R}\right)
\notag \\
\label{norm_stress_cont}
-\phi_L'\left[i\rho_L\left(kU_i-\omega\right)-\frac{\mu_L}{R^2}+3
\mu_Lk^2\right]
+\phi_L\left(i\rho_LkU_L'(R)+2\mu_L\frac{k^2}{R}\right)
\\
\label{stress_cont}
\mu_G\left[k^2\phi_G+\frac{kU_G''(R)}{\omega-kU_i}\phi_G+\phi_G''-\frac{\phi_G'}{R}\right]=
\mu_L\left[k^2\phi_L+\frac{kU_L''(R)}{\omega-kU_i}\phi_L+\phi_L''-\frac{\phi_L'}{R}\right]
\end{align}
Functions $\phi_G$ and $\phi_L$ correspond respectively to the total stream function in the gas and liquid phase; $\sigma$ is the liquid/gas surface tension.

The dispersion relation obtained from the previous system is solved for spatial branches for the experimental conditions presented in the following section: air and water at $T=20^{\circ}C$, with varying geometries ($R$ and $H_G$), gas/liquid velocities $U_G$ and $U_L$ and the corresponding measured vorticity thicknesses $\delta_G$ and $\delta_L$. Note that since liquid velocity is experimentally measured via the liquid flow rate, we enforce equality of the mean liquid velocity to $U_L$: this implies in particular that for $\delta_d=1$ and large $U_G$, the liquid velocity $U_{L0}$ far from the interface (see expression for the base flow) can be significantly smaller than $U_L$. 

For the conditions corresponding to our experiments, three situations can arise, corresponding to the different scenarios discussed in the introduction: the instability can be convective and driven by the Yih mode or absolute driven by a surface tension branch \citep{Otto}, or it can be absolute but with a pinch point at low wavenumber, between the shear and confinement branches \citep{Matas_confi}. The absolute instability is identified along the criteria introduced by \citet{Huerre}, in particular the imaginary part of the frequency has to be positive when the branches pinch, this is the absolute growth rate $\omega_{0i}$ ; the group velocity tends to zero at the pinch point ; in all the cases studied here the pinching occurs between the shear branch which lies in the positive $k_i$ half-plane at large $\omega_i$, and a branch (surface tension branch or confinement branch) which remains in the negative $k_i$ half-plane at large $\omega_i$.

\section{Experimental results and comparison to stability analysis}

We now present experimental results, and confront them to the predictions of the linear stability analysis. A choice must be made in the stability analysis regarding the shape of the base flow: as commented before, parameter $\delta_d$ determines if the profile is fully developed ($\delta_d$ close to 1) or presents a strong velocity deficit ($\delta_d\ll 1$). In the experiment $\delta_d$ is expected to be small close to the splitter plate, and to increase downstream, but the measurement of $\delta_d$ at finite distances is practically impossible due to the large growth rates of the instability once the gas flow is turned on.  We will therefore present for each geometry the predictions corresponding to several $\delta_d$, and discuss their respective plausibility and agreement with experimental data. 

\begin{figure}
\centering
\includegraphics[width=0.45\textwidth]{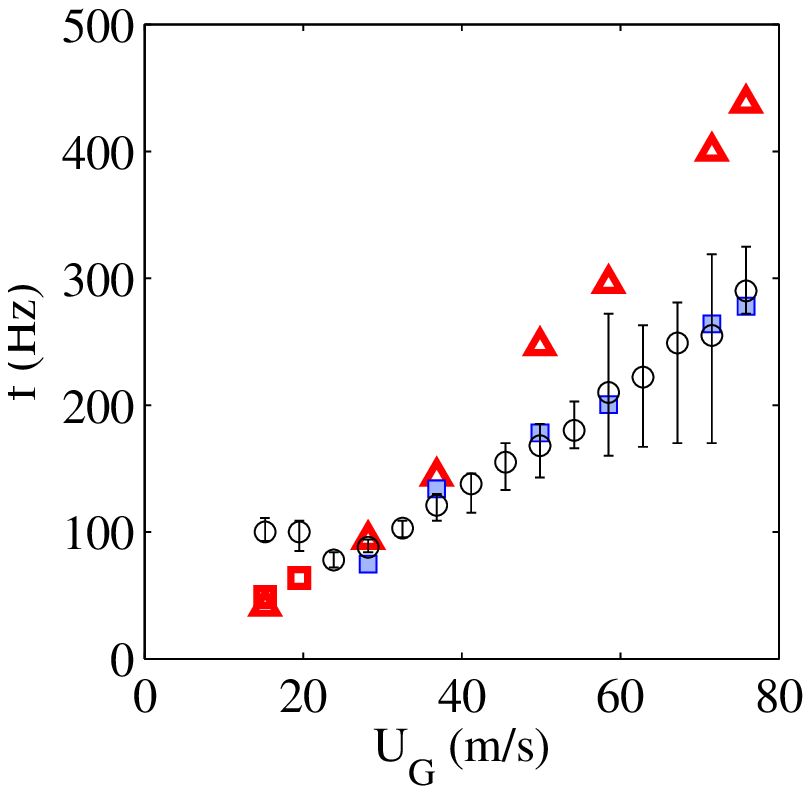}
\includegraphics[width=0.45\textwidth]{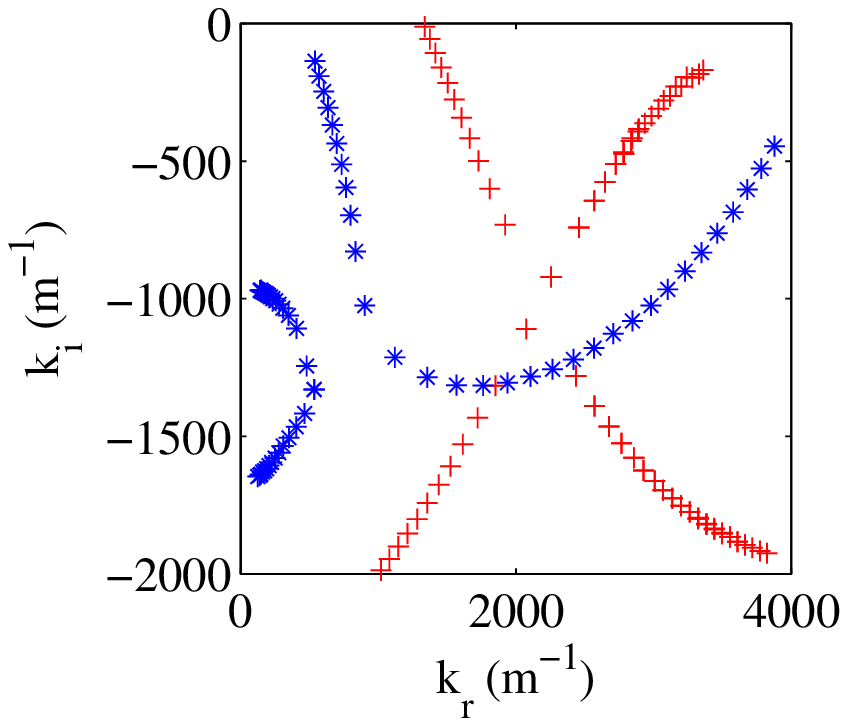}
\caption{Left: Frequency as a function of gas velocity, $H_G=5$~mm and $H_L=5$~mm, $U_L=0.28$~m/s: $\circ$ experimental results ; $\square$ viscous stability analysis prediction with $\delta_d=0.3$ ; $\triangle$ viscous stability analysis prediction with $\delta_d=0.1$. Red open symbols correspond to the surface tension absolute instability, blue filled symbols to the confinement absolute instability. Error bars on the experimental data indicate the half height width of the maximum peak in the spectrum. Right: spatial branches for same conditions and $U_G=37$~m/s, $\delta_d=0.3$ (${\color{blue}*}$) and $\delta_d=0.1$ (${\color{red}+}$), showing respectively the confinement ($\omega_{0i}=270~s^{-1}$) and surface tension ($\omega_{0i}=950~s^{-1}$) pinch points.}
\label{fig:Hg5Hl5}
\end{figure}

In figure \ref{fig:Hg5Hl5} we first show the comparison of experimental frequency (symbol $\circ$) to linear stability predictions for the geometry $H_G=5$~mm and $H_L=5$~mm, as a function of gas velocity, for a fixed $U_L=0.28$~m/s and for $\delta_d=0.3$ and $\delta_d=0.1$. The vorticity thickness in the liquid stream at $U_L=0.28$~m/s is deduced from similitude hot wire measurements in air flow at the relevant $Re$, and is $\delta_L=1\pm 0.1$~mm. The error bars on the experimental data indicate the half height width of the maximum peak in the spectrum: the larger error bars in the range $U_G = 60-75$ m/s are due to the non axisymmetric large scale instability discussed previously, which exhibits a very large growth rate for these conditions. We represent the destabilizing mechanism at play with the following color code: red open symbols correspond to the absolute surface tension mechanism, blue solid symbols to the absolute confinement mechanism. For this geometry and liquid velocity, the best agreement is found for $\delta_d=0.3$ (symbol $\square$). The mechanism is the surface tension absolute mechanism for the two lowest gas velocities, but it switches to the confinement mechanism for $U_G> 25$~m/s.

The prediction obtained with $\delta_d=0.1$ (symbol $\triangle$), which remains driven by the surface tension mechanism for all $U_G$, overestimates the experimental data, especially for larger gas velocities. Spatial branches around the pinch point are shown for both $\delta_d$ values and for $U_G=37$~m/s on figure \ref{fig:Hg5Hl5} right. The lower frequency associated with the confinement mechanism results from the fact that the confinement branch intercepts the shear branch at lower wavenumbers, hence at lower frequencies for this downstream propagating ($d\omega_r/dk_r>0$) shear mode. 

Values of $\delta_d$ larger than 0.3, i.e. larger interface velocities $U_i$, all lead to instability triggered with confinement: the frequency value decreases with increasing $\delta_d$, and since the $\delta_d=0.3$ assumption is in good agreement with the experimental series the $\delta_d>0.3$ values therefore underestimate the experimental values (not shown here). In addition, increasing $\delta_d$ leads to imposing interface velocities much larger than the liquid velocity $U_{L0}$, and this induces a difficulty in the consistency of the stability analysis with respect to the experimental situation: this leads to a much increased flow rate in the liquid boundary layer. In order to ensure that the overall liquid flow rate remains equal to the experimental one, and in order to mimick the experimental situation, one should therefore  reduce the radius at large gas velocities. Even though this may reflect the experimental situation where the liquid jet radius decreases along  downstream distance, and does this faster for larger $U_G$, we decided not to adjust geometrical parameters in the present study, and therefore limit ourselves to $\delta_d<0.3$ for the present case. This issue is inherently related to the assumption that a local stability analysis is valid, and that the base flow is a parallel flow, while in the experiments spatial variations (of radius or velocity deficit for example) certainly occur over distances smaller than the typical instability wavelength. At any rate, the aim here is to demonstrate that in spite of the strong parallel flow assumption the local stability analysis captures the experimental results and sheds light on the physical mechanisms at play.

\begin{figure}
\centering
\includegraphics[width=0.45\textwidth]{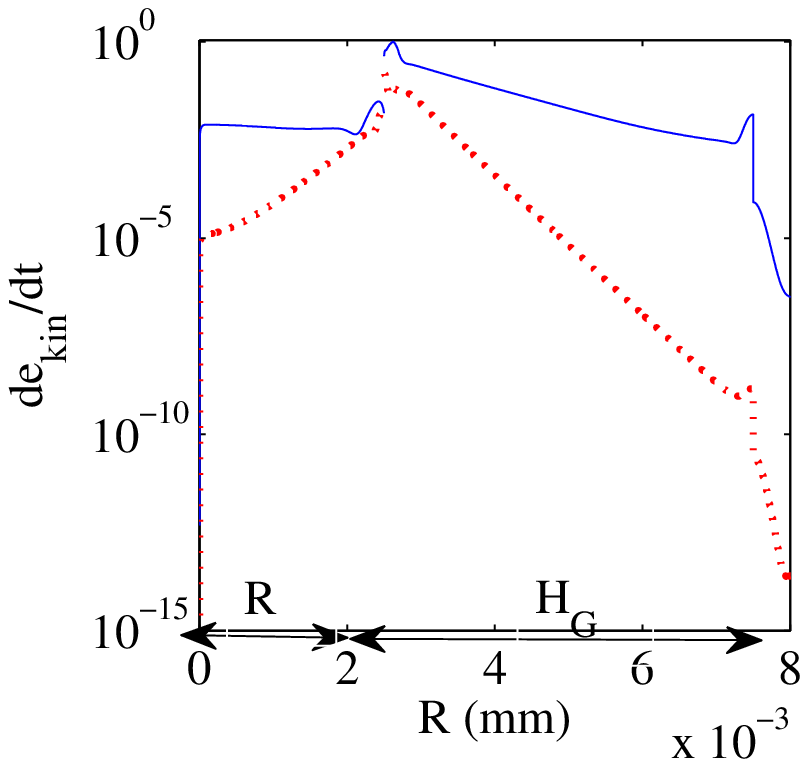}
\includegraphics[width=0.5\textwidth]{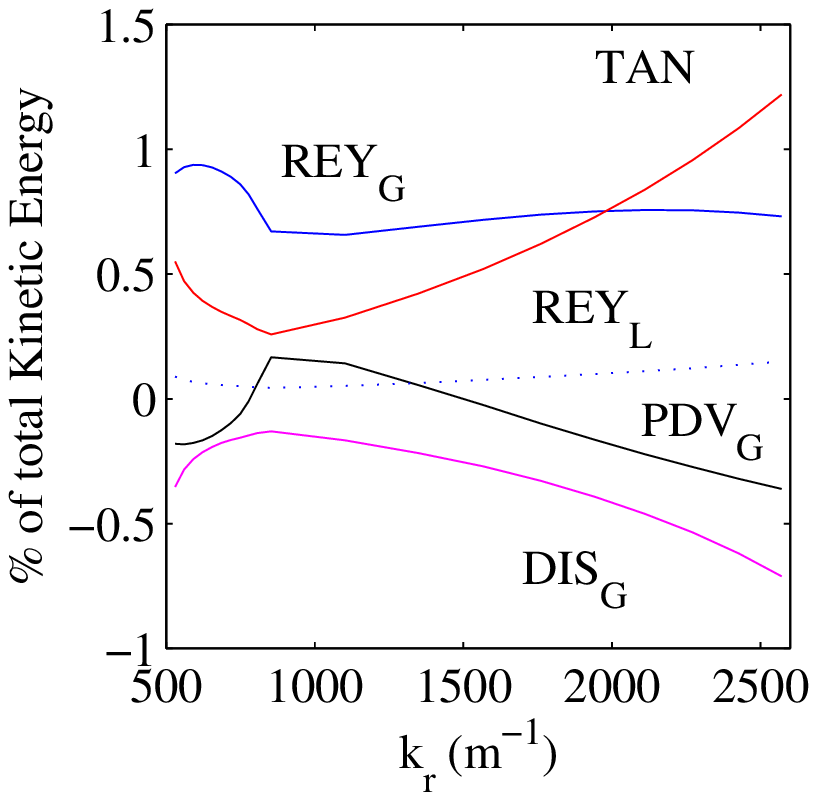}
\caption{Left: Comparison of the local rate of kinetic energy fed to the perturbation as a function of radial position for the conditions of the two pinch points of figure \ref{fig:Hg5Hl5} right, for $R=H_L/2=2.5$~mm and $H_G=5$~mm: the mode controlled by the surface tension mechanism (red dotted curve, $\delta_d=0.1$) is only active close to the interface, while the perturbation associated with the confinement mechanism (blue solid curve, $\delta_d=0.3$) fills the liquid and gas streams. Right: energy budget as a function of wavenumber for the shear mode of figure \ref{fig:Hg5Hl5} right, $\delta_d=0.3$. At low wavenumbers, and in particular close to the pinch point at $k_r\approx 1000$~m$^{-1}$, the velocity perturbation is mostly fed by gas Reynolds stresses.}
\label{fig:kin}
\end{figure}

The best agreement for the $H_G=5$~mm and $H_L=5$~mm geometry is therefore obtained when the instability is triggered by the confinement absolute instability. The origin and potential relevance of  confinement branches, which is explained in \citet{Healey2}, resides in the propagation of perturbations in the cross stream direction (presently the radial direction). The confinement branch of figure \ref{fig:Hg5Hl5} reaches its maximum $k_r$ for $k_i\approx 1200$~m$^{-1}$. This value corresponds to a length scale $L\sim 5$~mm, which is exactly the (equal) widths of the liquid and gas streams in this geometry. The physical idea is simply that when the shear branch pinches with the confinement branch the cross stream perturbations wavelength $2\pi/k_i$ matches the stream width: this resonance feeds the shear mode due to the negative group velocity of the confinement branch at the pinch point, and leads to the occurrence of the absolute instability. To illustrate further the perturbations associated with each pinching mechanism, we plot on figure \ref{fig:kin} left the variations of the rate of change of the kinetic energy density of the perturbation $de_{kin}/dt$ as a function of radial distance. This quantity is computed via the eigenfunction for the velocity perturbation, the formula corresponds to the integrand of the total kinetic energy rate per unit axial length $dE_{kin}/dt$ given in the appendix. This plot illustrates that the surface tension mechanism (red dotted curve) is only active close to the interface, while the perturbation associated with the confinement mechanism (blue solid line) fills the whole liquid and gas streams due to the cross stream resonance at low $k_r$.
In order to  analyze how the total kinetic energy rate per unit axial length $dE_{kin}/dt=\int (de_{kin}/dt)2\pi rdr$ of the perturbation is fed for given $k$ and $\omega$, we separate the contributions in the energy budget following the method introduced by \citet{Boomkamp}, and used by \citet{Otto} in a similar context. We note $REY_L$ and $REY_G$ the work of respectively liquid/gas Reynolds stresses, $DIS_L$ and $DIS_G$ the viscous dissipation in the liquid and in the gas, $TAN$ the work of tangential stresses at the interface and $NOR$ the work of normal stresses at the interface (surface tension). In the present case of non zero spatial growth rate, and therefore of upstream/downstream dissymetry, we must also consider the rate of work of pressure within the liquid and gas streams, which we note respectively $PDV_L$ and $PDV_G$. The expressions for each of these contributions are given in the appendix. The total budget then writes:
\begin{equation}
\frac{dE_{kin}}{dt}=REY_L+REY_G+PDV_L+PDV_G+DIS_L+DIS_G+TAN+NOR
\label{energy_budget}
\end{equation}
On figure \ref{fig:kin} right, we plot the variations of each contribution as a function of $k_r$, for the shear mode at $U_L=0.28$ m/s, $U_G=37$ m/s, $\delta_d=0.3$ (figure \ref{fig:Hg5Hl5} right, symbol ${\color{blue} *}$). 
The contributions are normalized by the left-hand side of equation (\ref{energy_budget}), the total rate at which kinetic energy increases. The shear branch passes close to the pinch point at $k_r\approx 1000$~m$^{-1}$: the dominant contribution for this wave number is $REY_G$, due to the gas Reynolds stresses.  It is interesting to note that close to the pinch point the $PDV_G$ contribution changes sign, and becomes positive: the fact that an ``overpressure'' in the gas stream feeds the perturbation for this $k_r$ can be seen as a signature of the confinement mechanism. The $PDV_L$, $DIS_L$ and $NOR$ contributions are not shown, they are negligible compared to the others. As expected, the viscous contribution $TAN$ becomes dominant at large wavenumbers.

\begin{figure}
\centering
\includegraphics[width=0.45\textwidth]{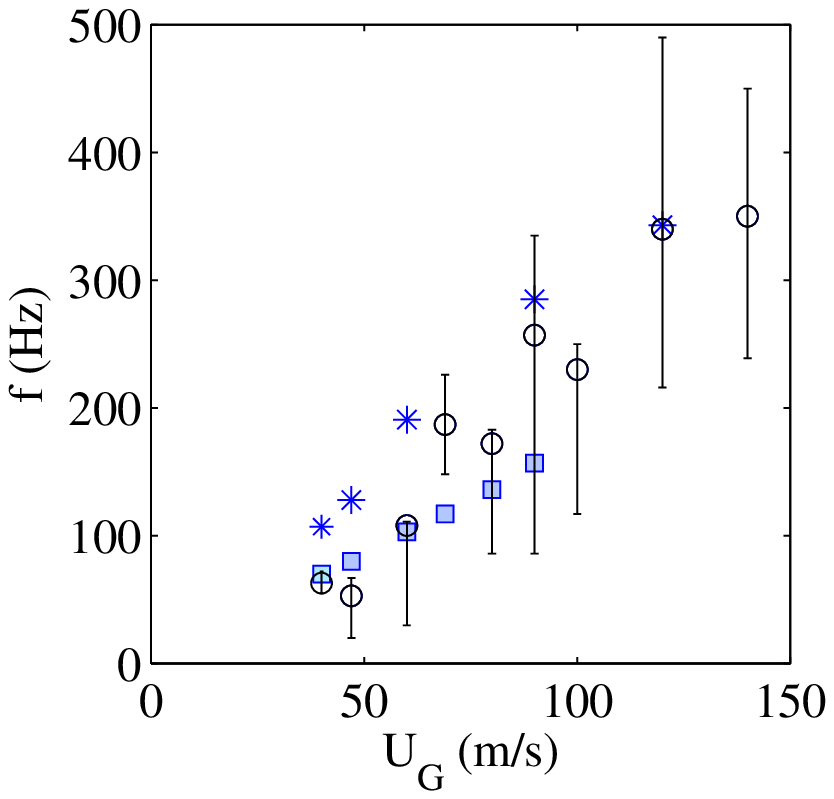}
\includegraphics[width=0.45\textwidth]{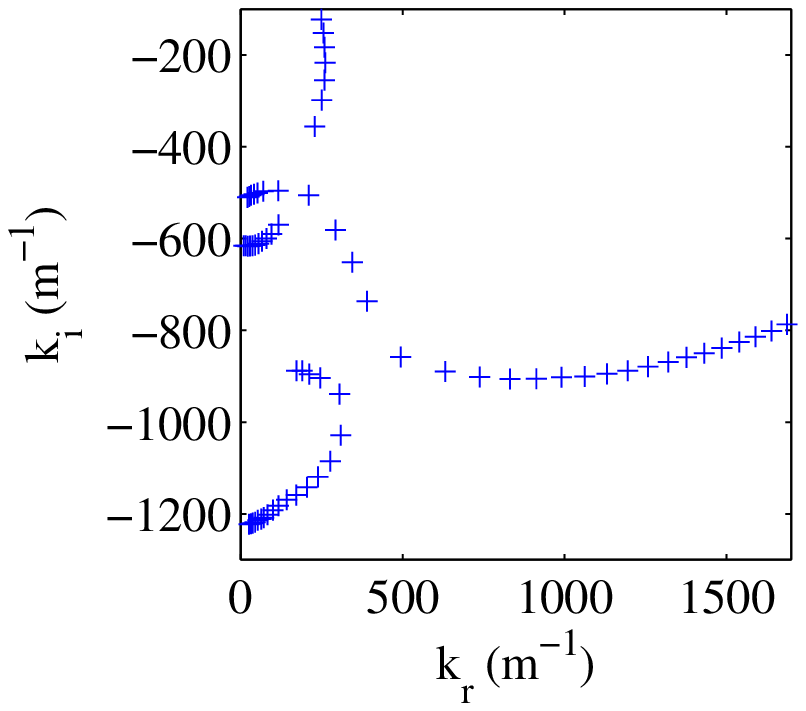}
\caption{Left: Frequency as a function of gas velocity, $H_G=5$~mm,  $H_L=20$~mm and $U_L=0.23$~m/s: $\circ$ experimental results ; {\color{blue}$\square$} viscous stability analysis prediction with $\delta_d=0.1$ ; {\color{blue}$*$} :  $\delta_d=0.075$. Error bars on the experimental data indicate the half height width of the maximum peak in the spectrum. Right: spatial branches for the $\delta_d = 0.1$ case, $U_G=47$~m/s, $\omega_{0i}=$210$~s^{-1}$.}
\label{fig:Hg5Hl20}
\end{figure}

In figure \ref{fig:Hg5Hl20} we now show the comparison of experimental frequency (symbol $\circ$) to linear stability prediction for the $H_G=5$~mm and $H_L=2R=20$~mm geometry, for a fixed $U_L=0.23$~m/s. Two values of $\delta_d$ are presented, $\delta_d=0.075$ and $\delta_d=0.1$. The smaller values of relevant $\delta_d$ compared to the previous geometry are due to the fact that parameter $\delta_d$ is made non dimensional with the liquid boundary layer $\delta_L$:  the much larger liquid jet radius in the present geometry leads to $\delta_L=5\pm 0.5$~mm for the considered liquid velocity, i.e. a much larger value than for the previous geometry. The length scales $\delta_d\delta_L$ associated with the relevant velocity deficits are similar in both cases, of the order of 300 - 500 $\mu$m. Under the assumption $\delta_d=0.1$, the maximum gas velocity compatible with $U_L=0.23$~m/s is $U_G=80$~m/s (meaning that $U_{L0}$ reaches zero above this value), larger $U_G$ are therefore not investigated for this series. We use the same color code as in figure \ref{fig:Hg5Hl5} to identify the destabilizing mechanism: the exclusively blue color of the symbols of figure \ref{fig:Hg5Hl20} means that for all these conditions the absolute instability occurs with a confinement branch. Figure \ref{fig:Hg5Hl20} right shows the pinch point for the case $U_G=47$~m/s and $\delta_d=0.1$ and the corresponding absolute growth rate $\omega_{0i}=210$~s$^{-1}$. The location of the confinement branch, at $k_i\sim 550~$m$^{-1}$, corresponds to a length $L\sim 1$~cm which is the radius of the liquid jet. Interestingly, a second confinement branch can be seen below, corresponding to half that length: a second pinch point would arise if $\omega_i$ were lowered below $\omega_{0i}=210$~$s^{-1}$. The corresponding frequency would be around $f=103~$Hz, slightly larger than the experimental value for this $U_G$. On figure \ref{fig:Hg5Hl20} left, only the frequency corresponding to the pinch point associated with the largest $\omega_{0i}$, and therefore the larger length $L$, has been retained for a given $\delta_d$. The strong impact of $\delta_d$ on the frequency is precisely due to the switching between these two confinement branches: larger frequencies associated with $\delta_d=0.075$ (symbol ${\color{blue} *}$) correspond to the branch around $k_i\approx 1100~$m$^{-1}$, while lower frequencies associated with $\delta_d=0.1$ (symbol ${\color{blue} \square}$) are all associated with pinching close to $k_i\approx 550~$m$^{-1}$. The existence of these two competing mechanisms in a narrow $\delta_d$ range may explain the non monotonous behaviour and large error bars in our measurements for this particular series.

\begin{figure}
\centering
\includegraphics[width=0.45\textwidth]{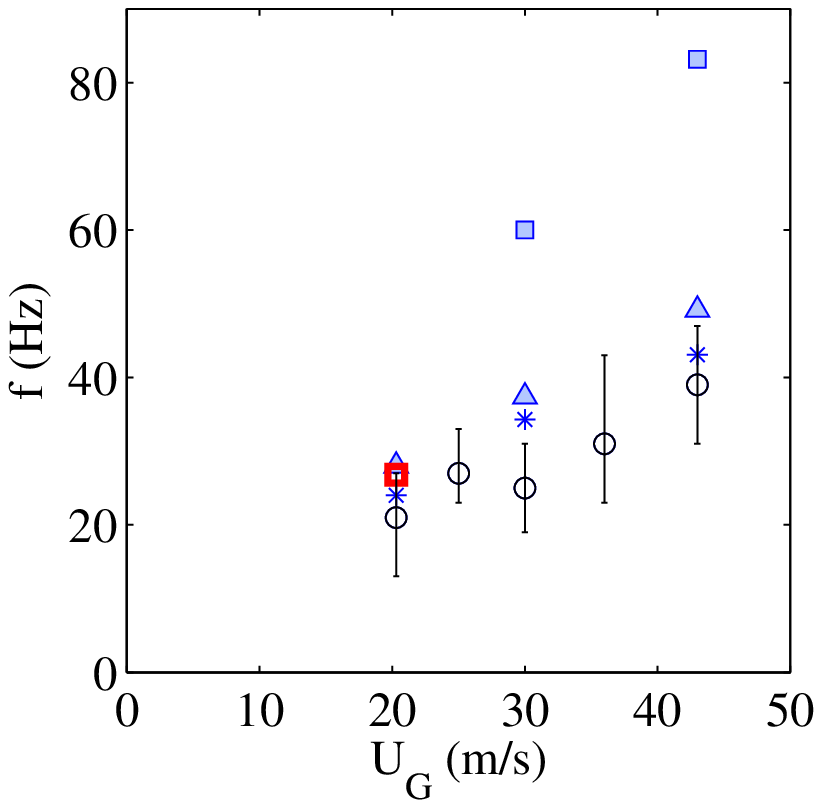}
\includegraphics[width=0.45\textwidth]{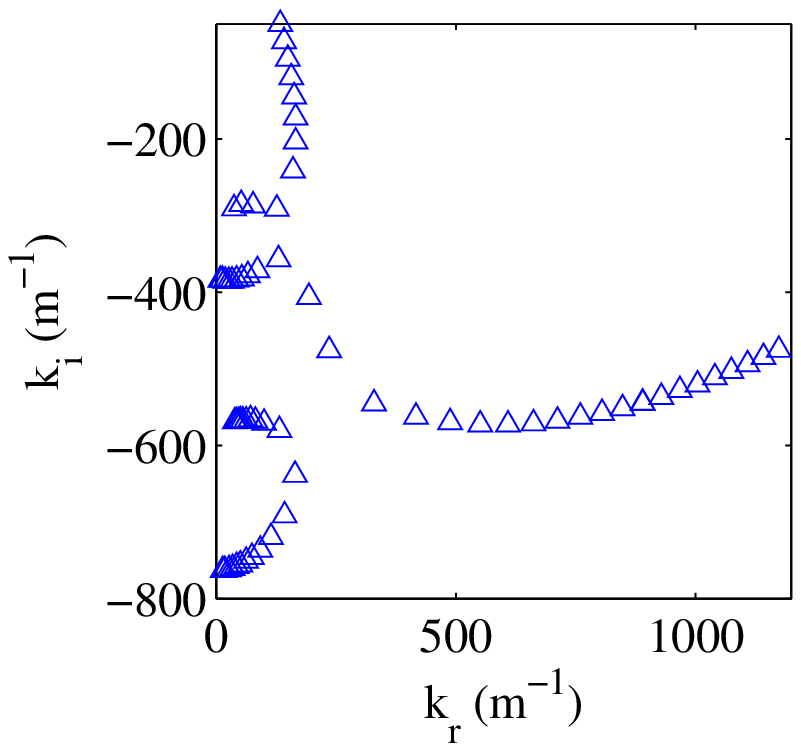}
\caption{Left: Frequency as a function of gas velocity, $H_G=24$~mm,  $H_L=20$~mm and $U_L=0.23$~m/s: $\circ$ experimental results ; $*$ viscous stability analysis prediction with $\delta_d=0.25$ ; $\bigtriangleup$  $\delta_d=0.2$ ; $\square$  $\delta_d=0.1$.  Error bars on the experimental data indicate the half height width of the maximum peak in the spectrum. Right: spatial branches for the $\delta_d = 0.2$ case, $U_G=43$~m/s and $\omega_{0i}=90~s^{-1}$.}
\label{fig:Hg24Hl20}
\end{figure}

Finally, figure \ref{fig:Hg24Hl20} shows the experimental frequencies for the $H_L=20~$mm and $H_G=24$~mm geometry, for a fixed $U_L=0.23$~m/s (symbol $\circ$). The agreement is best for $\delta_d=0.25$, symbol {\color{blue} *}, which corresponds to the confinement mechanism.   The pinch point arises because of confinement for all conditions except for the lowest $U_G$ of the $\delta_d=0.1$ assumption, where it occurs because of surface tension. As observed in the $H_G=5$~mm and $H_L=2R=5$~mm geometry, the surface tension mechanism becomes dominant for the lowest $\delta_d$ value, corresponding to the lowest interface velocity, but the frequency predicted with $\delta_d=0.1$ largely overestimates the experimental data for $U_G> 30$~m/s. Figure \ref{fig:Hg24Hl20} right shows the spatial branches around the pinch point for $\delta_d = 0.2$ and $U_G=43$~m/s. The shear branch pinches with a confinement branch located around $k_i\sim 350~$m$^{-1}$: this corresponds to a length $L\sim 2~$cm, which corresponds roughly to the  gas channel width $H_G$. As for the previous geometry, a higher order confinement mode can be seen below.

The comparison of the experimental frequencies for the $H_L=5~$mm-$H_G=5$~mm and $H_L=20~$mm-$H_G=24$~mm geometries shows that at similar liquid and gas velocities, the frequencies are three times as small in the larger geometry. We will discuss in the next section how geometry directly impacts frequency.

\begin{figure}
\centering
\includegraphics[width=0.45\textwidth]{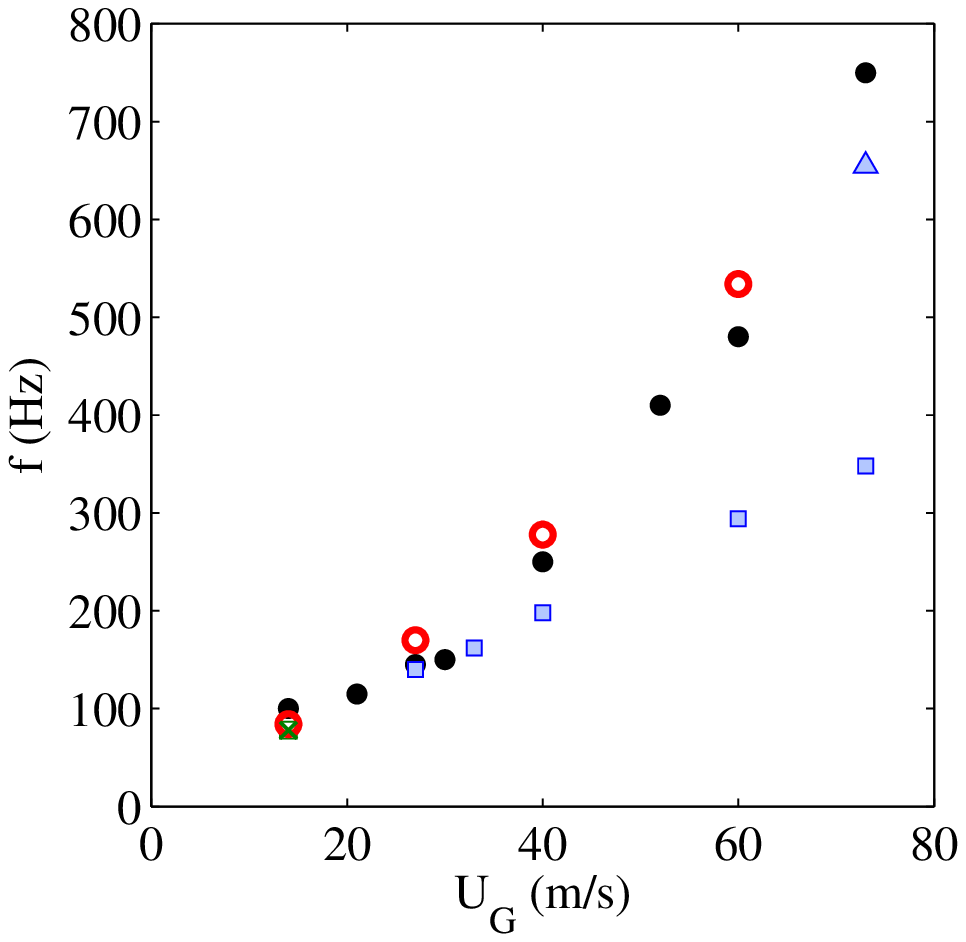}
\includegraphics[width=0.45\textwidth]{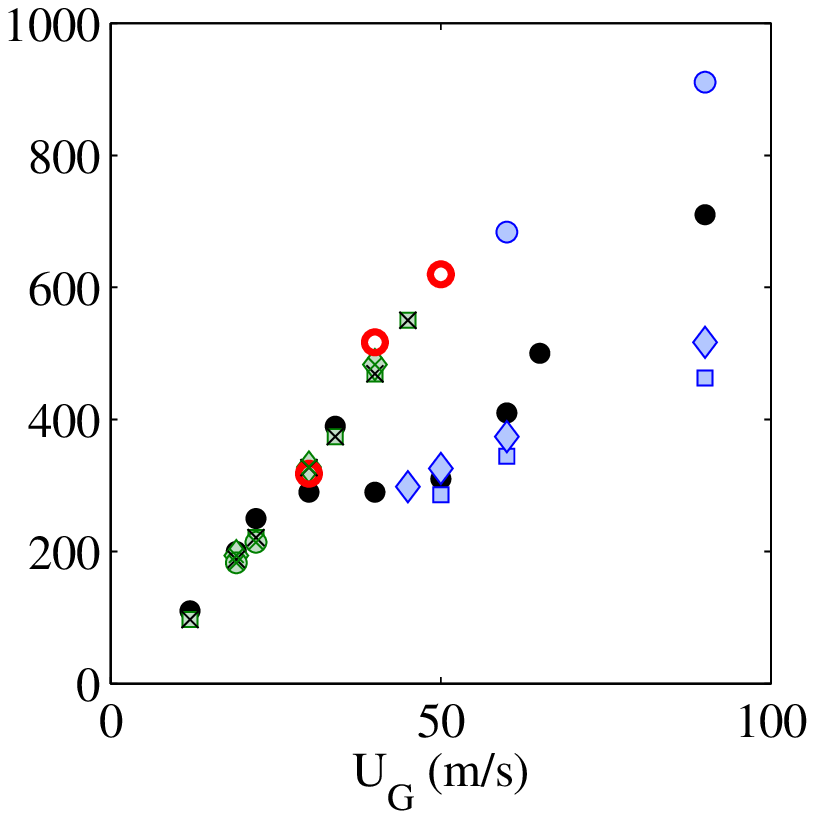}
\caption{Frequency as a function of gas velocity, $H_G=1.7$~mm and $H_L=8$~mm. Left: $U_L=0.45$~m/s ; $\bullet$ experimental results of Marmottant \& Villermaux (2004) \cite{Marmottant} ; $\square$ viscous stability analysis prediction with $\delta_d=1$ ; $\circ$ viscous stability analysis prediction with $\delta_d=0.5$ ; Right: $U_L=0.8$~m/s. Same symbols and $\lozenge$ $\delta_d=0.7$. Green symbols correspond to a convective instability, red symbols to the surface tension absolute mechanism, and blue symbols to absolute instability controlled by confinement. }
\label{fig:Marmottant}
\end{figure}

In figure \ref{fig:Marmottant} we compare the experimental frequency measured by \citet{Marmottant} on an analogous coaxial jet geometry as a function of gas velocity $U_G$ to predictions of the present stability analysis, for two fixed mean liquid velocities. As for previous geometries, we use different symbols to represent different values of $\delta_d$ in the base flow velocity profile injected into the analysis. The liquid boundary layer is estimated at $\delta_L=200~\mu$m: this quantity is not measured in \citet{Marmottant}, but we follow \citet{Otto} in estimating that it is close to $\delta_G$. It is at any rate expected to be much smaller than in the previous geometries due to the strong convergence ratio of the nozzle \citep{Marmottant}. The same color code as before is used, and we now put a cross within open green symbols to indicate conditions for which the instability is convective (e.g. {\color{green} $\boxtimes$}). For the case $U_L=0.45$~m/s (figure \ref{fig:Marmottant} left), the best agreement is found for $\delta_d=0.5$: the mechanism is the surface tension mechanism for all gas velocities (as indicated by the red color of symbols), except for the largest one $U_G=73$~m/s. For this gas velocity the instability is absolute but with a confinement branch of order two (such as in figure \ref{fig:Hg24Hl20} right for $k_i\sim -650~$m$^{-1}$). For the case $U_L=0.8$~m/s (figure \ref{fig:Marmottant} right), the non monotonous variations of the experimental frequency (symbol $\bullet$) suggest that a change in the mechanism occurs between $U_G=35$~m/s and $U_G=40$~m/s. This change is well captured by the stability analysis, by both the $\delta_d=0.7$ and $\delta_d=1$ assumptions. In both cases, the instability is convective for lower gas velocities (green symbols): the frequency is then almost independent of $\delta_d$. However,  it switches to an absolute instability driven by confinement when $U_G$ is increased: for $U_G>40$~m/s when $\delta_d=0.7$, and above $U_G=45$~m/s for $\delta_d=1$. This confinement mechanism occurs around a confinement branch located at $k_i \sim 1500~$m$^{-1}$, corresponding to $L\sim 4$~mm which is the value of the liquid jet radius. The assumption $\delta_d=0.5$ (symbol $\circ$), associated to a smaller interface velocity, leads to a dominance of the surface tension mechanism until $U_G=60~$m/s, and therefore to an overestimation of the experimental frequency.  Above $U_G=60$~m/s,  the shear branch pinches with a second order confinement branch.





\section{Discussion}

The  results of the previous section have illustrated how different competing mechanisms can become relevant when the width or velocity of the gas and liquid streams are modified. We now wish to discuss the scaling laws relevant in each regime, as well as criteria for predicting the occurrence of each of these mechanisms, and confront these criteria to the previous results.

The two possible absolute instability mechanisms occur when the shear branch pinches with either a confinement or  surface tension branch. In order to capture the frontier between the different regimes, it is therefore necessary to describe first how this shear mode scales. As explained in \citet{Otto} when the air/water surface tension is taken into account, a cut-off at large $k_r$ is introduced, and the initially distinct inviscid and viscous modes merge. The location of the most dangerous mode wavenumber $k_{r\;max}$ corresponds to a balance between surface tension (which controls the cut-off) and the destabilizing mechanism. We follow \citet{Hinch} and write that surface tension will take over for $We_{\gamma}=\frac{\rho_G U_G^2}{k^3\delta_G^2\sigma} < 1$. This condition can be derived within the context of a purely temporal approach by simplifying the balance of normal stresses at the interface as $\mu_G \zeta_G$ (destabilizing) versus $\sigma k^2 \eta_0$ (stabilizing), where $\zeta_G$ is the vorticity perturbation generated in the gas stream by the interface perturbation $\eta_0$. By writing that $\zeta_G=(U_G/\delta_G)^2\eta_0/(\nu_Gk)$ \citep{Hinch}, the ratio of these two normal stress contributions simplifies into the expression for $We_{\gamma}$. The wavenumber $k_{r\;max}$ corresponding to the maximum growth rate is taken as the one for which $We_{\gamma}\sim 1$:
\begin{equation}
k_{r\;max} \sim \left(\frac{\rho_G}{\sigma}\right)^{1/3}\left(\frac{U_G}{\delta_G}\right)^{2/3}
\label{eq:kr}
\end{equation}

\begin{figure}
\centering
\includegraphics[width=0.45\textwidth]{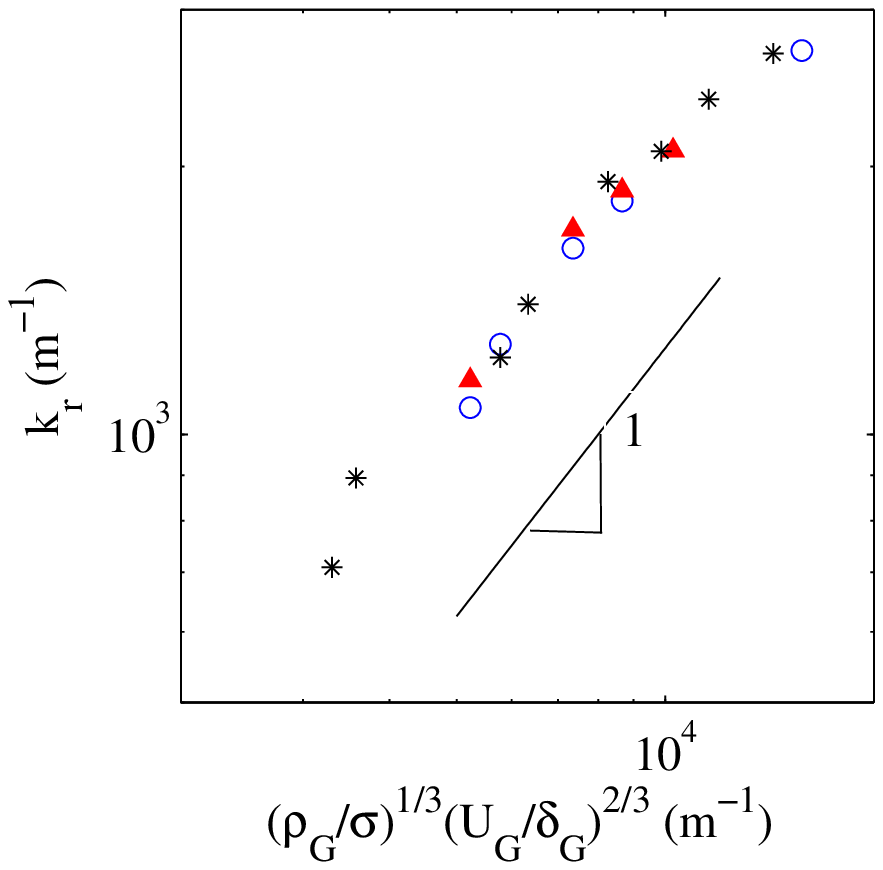}
\includegraphics[width=0.45\textwidth]{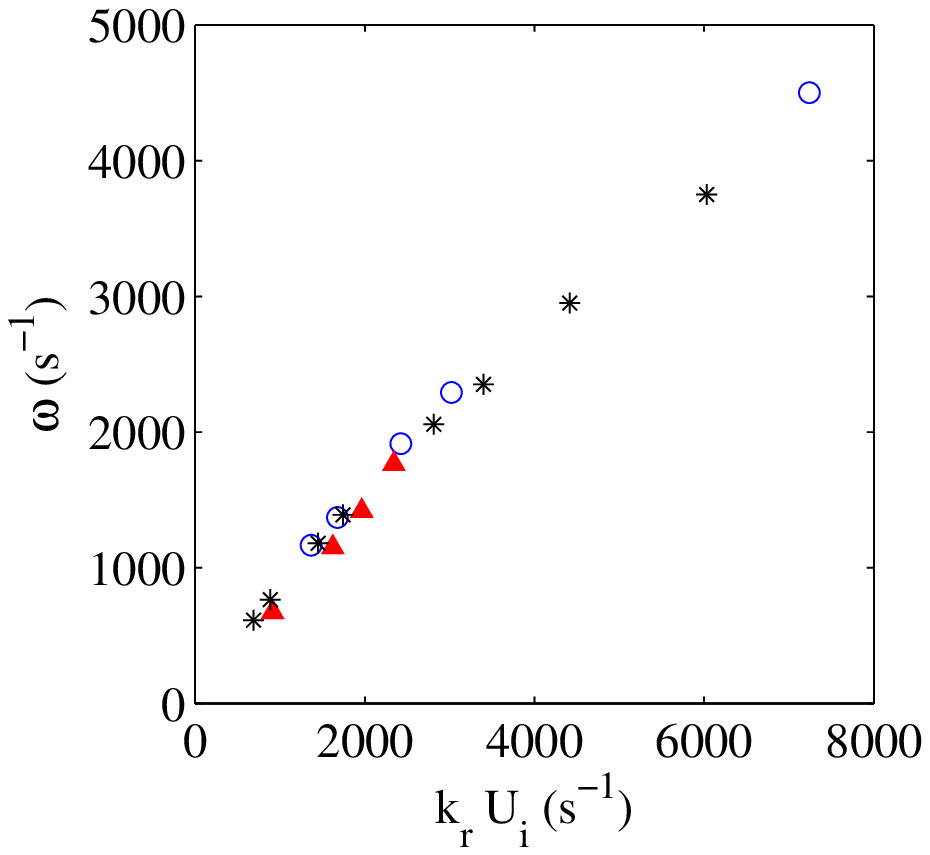}
\caption{Left: predicted wavenumber of the most dangerous mode for the injector of \citet{Marmottant}, for three liquid velocities: {\color{red} $\blacktriangle$} $U_L=0.45$~m/s, * $U_L=0.8$~m/s,  {\color{blue}$\circ$} $U_L=0.94$~m/s, gas velocity in the range 12-60~m.s$^{-1}$. These points include the convective conditions of figure \ref{fig:Marmottant}. Right: predicted pulsation $\omega$ as a function of $k_rU_i$.}
\label{fig:scaling_convectif}
\end{figure}

Figure \ref{fig:scaling_convectif} shows the comparison of this expression with the wavenumber of the most unstable (convective) mode predicted for the injector of figure \ref{fig:Marmottant}, for three liquid velocities and various $U_G$ and $\delta_G$ values. The dependence in the shear rate corresponds to the prediction of the analysis, even though to fit the stability analysis prediction the threshold of  $We_{\gamma}$ has to be set at $We_{\gamma}\approx 10^2$: The expression for $k_{r\;max}$ in equation (\ref{eq:kr}) should therefore be corrected with a coefficient $10^{-2/3}$ to fit the stability analysis data. Note also that though $We_{\gamma}$ has been introduced by \citet{Hinch} within a temporal analysis assumption, the comparison of figure \ref{fig:scaling_convectif} shows that it remains a relevant dimensionless number in our present spatial analysis.

Though this shear mode is expected to be a viscous one, viscosity does not appear in  equation (\ref{eq:kr}): the balance of normal stresses discussed above should actually be corrected to also include the contribution of $\zeta_L$, the vorticity perturbation on the liquid side. The balance then writes:
$$\mu_G \zeta_G+\mu_L \zeta_L=\sigma k^2\eta_0$$
$$\rho_G \frac{U_G^2}{\delta_G^2k}\eta_0+\rho_L \frac{(U_i-U_L)^2}{\delta_L^2k}\eta_0=\sigma k^2\eta_0$$
Injecting equation (\ref{eq:Ui}) for the interface velocity, this simplifies into:
$$\rho_G \frac{U_G^2}{\delta_G^2}+\rho_L \frac{U_G^2}{\delta_G^2}\frac{\mu_G^2}{\mu_L^2}=\sigma k^3$$
\begin{equation}
1+\frac{\rho_L}{\rho_G}\frac{\mu_G^2}{\mu_L^2}=\frac{\sigma k^3 \delta_G}{\rho_GU_G^2}=We_{\gamma}^{-1}
\label{eq:mu}
\end{equation}
For air and water, the corrective term in the left-hand side of equation (\ref{eq:mu}) is of the order of 0.2, and this shows that the destabilization is mostly caused by the vorticity perturbation on the gas side as initially assumed. In this limit, the most unstable wavenumber is expected to be independent of viscosity. If the viscosity ratio increases (if a less viscous liquid is used for example), the instability could be dominated by the shear on the liquid side, even though the shear rate is higher in the gas stream.

\begin{figure}
\centering
\includegraphics[width=0.45\textwidth]{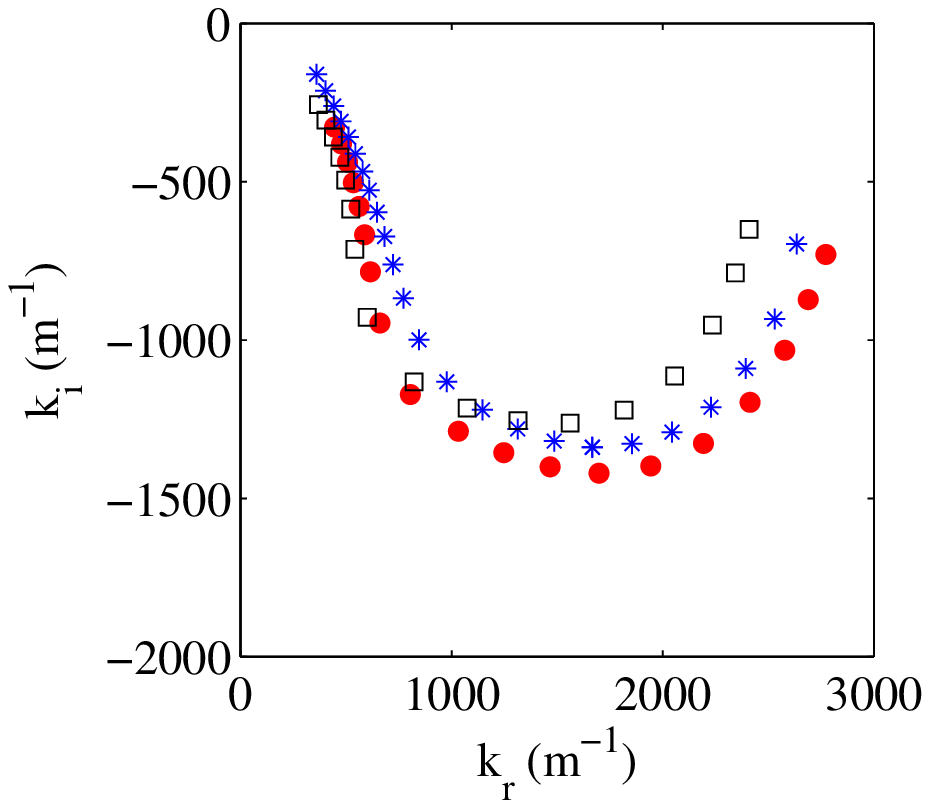}
\includegraphics[width=0.45\textwidth]{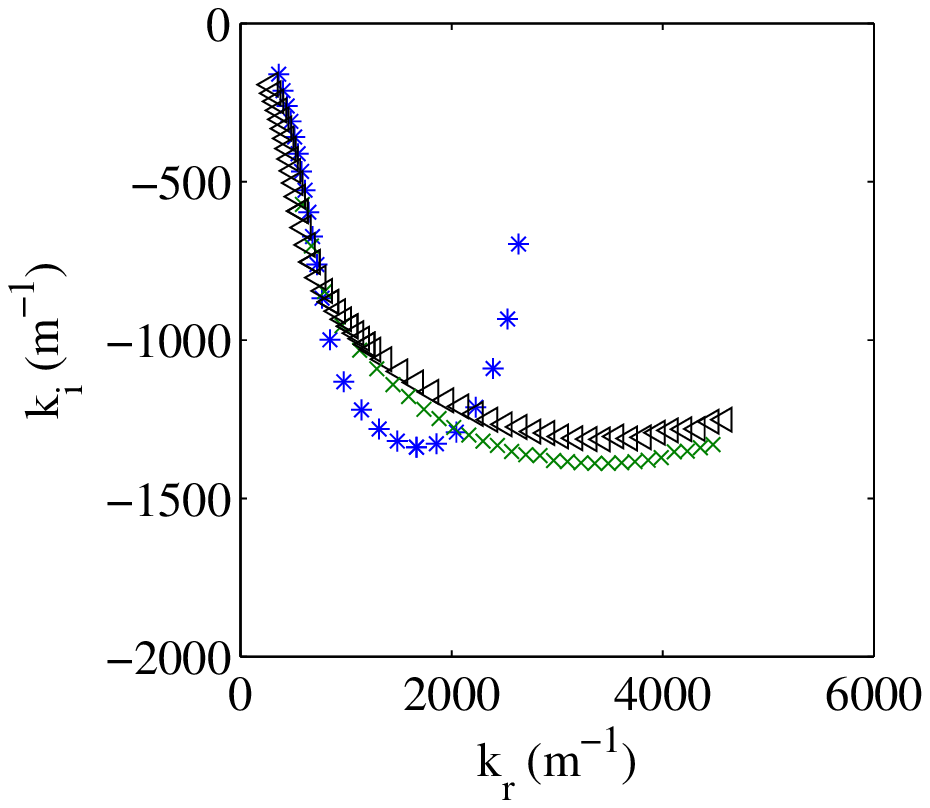}
\caption{Left: Illustration of the weak impact of  liquid and gas viscosities on the shear branch for $U_L=0.45$~m/s, $U_G=28$~m/s, and fixed $\nu_G/\nu_L$: {\color{red} $\bullet$} $\nu_G=\nu_{air}$ and $\nu_L=\nu_{water}$ ; {\color{blue} *} $\nu_G=4\nu_{air}$ and $\nu_L=4\nu_{water}$ ; $\square$ $\nu_G=\nu_{air}/4$ and $\nu_L=\nu_{water}/4$. Densities $\rho_L$ and $\rho_G$ are kept constant, equal to air/water values. Right:  An increase of the viscosity ratio from air/water values increases the wave number, due to the stronger impact of vorticity in the liquid stream. $\triangleleft$ \; $\nu_G=\nu_{air}$ and $\nu_L=\nu_{water}/4$ ; {\color{olive} $\times$} $\nu_G=4\nu_{air}$ and $\nu_L=\nu_{water}$.}
\label{fig:inf_nu}
\end{figure}

The limited impact of viscosity in the air/water case is illustrated in  figure \ref{fig:inf_nu} left for a fixed $U_G=28$~m/s and $U_L=0.45$~m/s (second point in the series $U_L=0.45$~m/s, figure \ref{fig:Marmottant}):  increasing or decreasing both gas and liquid viscosities by a factor of four, at fixed $\nu_G/\nu_L$ ratio and fixed $\rho_G$ and $\rho_L$, only induces a weak variation of $k_{r\;max}$: 2\% increase in $k_r$ if both viscosities are increased, 10\% decrease if both are divided by four. The only way viscosity can affect the shear mode is if the viscosity ratio $\mu_G/\mu_L$ is increased (see equation \ref{eq:mu}). This is illustrated in figure \ref{fig:inf_nu} right. Note finally that there is no effect whatsoever of viscosity on the maximum growth rate of the shear branch (the expression of the growth rate will be discussed below), even in the case of figure \ref{fig:inf_nu} right.

Conversely, dividing surface tension by four for the same conditions $U_G=28$~m/s and $U_L=0.45$~m/s (down to 18 mN/m) leads to an increase of $k_r$ of about 30\%, confirming that surface tension plays a role in mode selection for these conditions. However, absolute instabilities quickly take over if $\sigma$ is increased due to its impact on $k_r$ and $k_i$. Our interest here being to discuss  the scaling laws for the shear branch in the convective regime, a more systematic variation of $\sigma$ is not carried out.

Frequency can be estimated from the wavenumber by assuming that for the most unstable wavelength the instability travels at the interface velocity $U_i$. Figure \ref{fig:scaling_convectif} right  shows that the frequency of the most unstable mode can be estimated correctly by $k_rU_i$, and hence by $\left(\rho_G/\sigma\right)^{1/3}\left(U_G/\delta_G\right)^{2/3}U_i$. Note that though frequency does depend on liquid velocity (compare $U_L=0.45$~m/s and $U_L=0.8$~m/s series in figure \ref{fig:Marmottant}), all data are collapsed when rescaled via $U_i$. This expression predicts that when $\delta_G \propto U_G^{-1/2}$ frequency will increase as $f \propto U_GU_i$ in the convective regime: at low gas velocities $U_i$ is weakly dependent on $U_G$ and the scaling reduces to $f\propto U_G$. This is consistent with the experimental data of the $U_L=0.8$~m/s series in figure \ref{fig:Marmottant} right, as long as the predicted mechanism is the convective one (green crossed symbols).

We now turn to the spatial growth rate of the most unstable mode $k_{i\;conv}$, which is crucial since it will determine the location of the shear branch relative to the confinement/surface tension branches discussed below. 
 For low $k_r$, we consider that surface tension is negligible, and that the previously introduced pressure perturbation at the interface  will be balanced by liquid inertia. However, at large wavelengths the expression introduced earlier for the pressure perturbation cannot be valid anymore: due to the finite thickness of the boundary layer on the gas side, this pressure must saturate at $\mu_G \zeta_G \sim \rho_G U_G^2k_r \eta_0$ when $k\delta_G \ll 1$. This can be derived via the vorticity equation exactly as in \citet{Hinch}, but by considering that diffusion of vorticity is controlled by the shorter scale $\delta_G$ rather than by $k^{-1}$.
Inertia can be estimated as $\rho_L (k_iU_i)^2\eta_0$ where $(k_iU_i)^{-1}$ is the time scale associated to the perturbation growth. The balance between inertia and pressure then writes:
$$\rho_L k_i^2U_i^2\eta_0 = \rho_G U_G^2k_r^2 \eta_0+\rho_L\left(\frac{\delta_L\mu_G}{\delta_G\mu_L}\right)^2 U_G^2k_r^2 \eta_0$$
$$ k_i^2U_i^2 = \frac{\rho_G}{\rho_L} U_G^2k_r^2 +\left(\frac{\delta_L\mu_G}{\delta_G\mu_L}\right)^2 U_G^2k_r^2 $$
\begin{equation}
k_i=-\sqrt{\frac{\rho_G}{\rho_L}+\left(\frac{\delta_L\mu_G}{\delta_G\mu_L}\right)^2}\frac{U_G}{U_i}k_r
\label{eq:ki_complet}
\end{equation}
We assume that $\delta_G$ is not too small compared to $\delta_L$, in this case the liquid contribution can be neglected and the previous equation simplifies into: 
\begin{equation}
k_i=-\sqrt{\frac{\rho_G}{\rho_L}}\frac{U_G}{U_i}k_r
\label{eq:ki}
\end{equation}
We can now estimate $k_{i\;conv}$ from the value of $k_{r\;max}$ (equation \ref{eq:kr}), which we inject into equation (\ref{eq:ki}).
\begin{figure}
\centering
\includegraphics[width=0.45\textwidth]{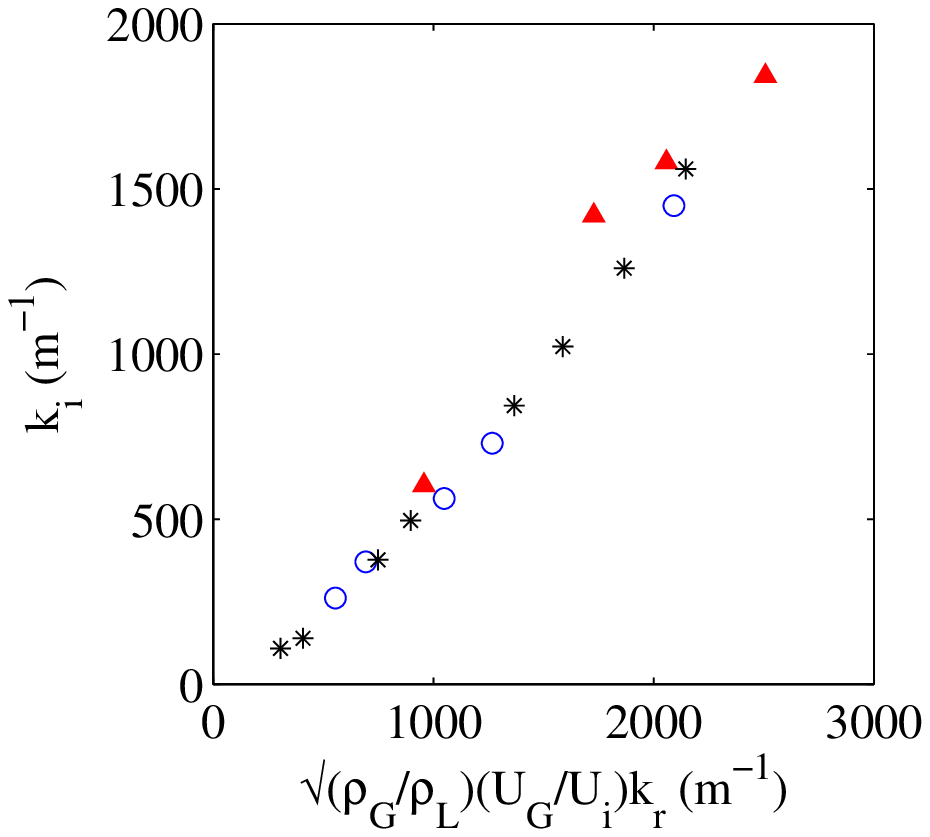}
\includegraphics[width=0.45\textwidth]{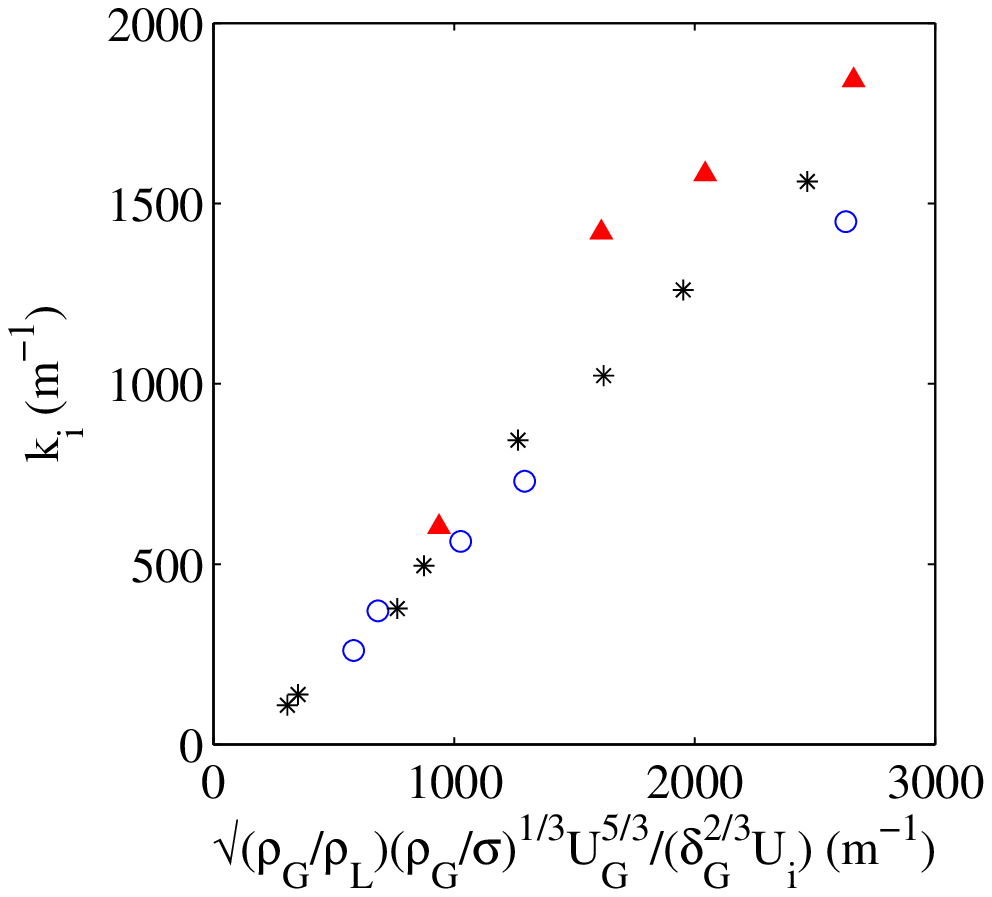}
\caption{Left: predicted spatial growth rate of the most dangerous mode as a function of $\sqrt{\rho_G/\rho_L}(U_G/U_i)k_{r}$, same symbols as figure \ref{fig:scaling_convectif}. Right: same plot, with $k_{r\;max}$ estimated via equation (\ref{eq:kr}).}
\label{fig:scaling_convectif2}
\end{figure}
Figure \ref{fig:scaling_convectif2} left shows that equation (\ref{eq:ki}) is largely valid for the stability analysis data of figure \ref{fig:Marmottant} where the convective instability is observed. Note that for this data the maximum growth rate $k_{i\;conv}$ depends on $U_G$ but also depends significantly on the liquid velocity: the impact of $U_L$ on the growth rate is fully captured through the role of $U_i$ in equation (\ref{eq:ki}), since all series are collapsed on a single curve. Finally, we had commented earlier the fact that in figure \ref{fig:inf_nu} right multiplying the viscosity ratio by four almost doubled the wave number but left the spatial growth rate of the most dangerous mode unchanged: the present model predicts this, since in equation (\ref{eq:ki}) the doubling of $k_{r\;max}$ is compensated by a doubling of the interfacial velocity when $\mu_G/\mu_L$ is multiplied by four for these conditions.

Figure \ref{fig:scaling_convectif2} right compares the model $k_{i\;conv}$  obtained when $k_{r\;max}$ is expressed directly with equation (\ref{eq:kr}) to the growth rate of the most dangerous mode predicted by the viscous stability analysis. The agreement is rather satisfying, given the series of assumptions made in deriving this expression. In particular, the strong effect of liquid velocity on $k_{i\;conv}$ is captured by the model. This is less true for the largest $k_{i\;conv}$ investigated: these points correspond to conditions for which the Weber number becomes larger than 200, and for which equation (\ref{eq:kr}) for the wavenumber does not hold anymore: the surface tension cut-off occurs at a wavenumber larger than the most dangerous mode at $k_{r\;max}=\delta_L^{-1}$. The shear rate above which this change of regime occurs is found by writing $We_\gamma>1$ for $k=\delta_L^{-1}$: $U_G/\delta_G>\sqrt{\sigma/\rho_G}\delta_L^{-3/2}$.

With these estimates in mind, we can now better understand in which conditions the absolute instability driven by confinement takes over the convective regime: this mechanism takes place when the slope of the $k_i(k_r)$ branch at low $k_r$ is such that the shear branch approaches the imaginary axis enough to pinch with the confinement branch. The pinching will typically occur at $k_i\sim k_r\sim 1/L$. The confinement induced absolute instability will therefore occur when the slope of the $k_i(k_r)$ branch is larger than one, i.e. when 
\begin{equation}
\sqrt{\frac{\rho_G}{\rho_L}}U_G>U_i
\label{eq:conf}
\end{equation}
This equation can be rewritten in terms of the dynamic pressure ratio $M=\frac{\rho_GU_G^2}{\rho_GU_L^2}$:
$$\sqrt{M}>1+\frac{\mu_G}{\mu_L}\frac{\delta_L}{\delta_G}\frac{U_G}{U_L}$$
$$\sqrt{M}\left(1-\frac{\mu_G}{\mu_L}\frac{\delta_L}{\delta_G}\sqrt{\frac{\rho_L}{\rho_G}} \right)>1$$
For air and water at ambient conditions, this condition simplifies into:
\begin{equation}
\sqrt{M}\left(1-0.5\frac{\delta_L}{\delta_G} \right)>1
\label{eq:conf2}
\end{equation}
If both boundary layers are of the same order of magnitude, this predicts that the mechanism will be convective at low $M$, and absolute due to confinement at large $M$. There are several restrictions to the preceding result: 
\begin{itemize}
\item[i)] The simple expression of equation (\ref{eq:conf2}) has been derived for the case where there is no velocity deficit in the base flow ($\delta_d=1$). A velocity deficit ($\delta_d<1$) is expected to favor the absolute confinement mechanism compared to the convective one, by decreasing the interfacial velocity. In addition, introducing a velocity deficit potentially modifies the sign and intensity of the vorticity perturbation $\zeta_L$ in the liquid boundary layer (as already evoked in deriving equation \ref{eq:mu}), and we will not attempt to quantify this in this work. 
\item[ii)] More generally, the liquid vorticity perturbation has been neglected in deriving equation (\ref{eq:conf2}). The idea here is to describe the onset of the confinement mechanism at moderate gas velocities, and condition (\ref{eq:conf}) is expected to be valid in such conditions. At larger gas velocities, and hence smaller $\delta_G$, the liquid contribution proportional to $U_i\propto U_G/\delta_G$ will eventually dominate, namely if $\delta_G \ll \delta_L(\mu_G/\mu_L)\sqrt{\rho_L/\rho_G}$: in these conditions it can be shown that $k_i\sim -k_r$ in the limit $k\delta_G \ll 1$, and hence that the confinement mechanism remains relevant at large $U_G$.  
\item[iii)] Equation (\ref{eq:ki}) has been derived for $\omega_i=0$: when $\omega_i>0$, this equation must be modified as $k_i\sim -\sqrt{\rho_G/\rho_L}(U_G/U_i)k_r+\omega_i/U_i$, and this relation can be inversed to provide an estimate of the absolute growth rate for the confinement absolute instability: $\omega_{i0} \sim \sqrt{\rho_G/\rho_L}U_G/L-U_i/L$. For the reasons mentioned above this estimate only holds for $\delta_d=1$. 
\end{itemize}
As a final remark, we also note that there is an additional obvious necessary condition for the confinement mechanism to take place: the smallest spatial growth rate $k_{i\;max}$ of the shear branch (its lowest point) must reach the location of the confinement branch, i.e. $k_{i\;max}L>1$. This means that at fixed velocities, the system must be large enough so that the confinement branch is within reach. Conversely, for a given stream width, $k_{i\;max}$ (or equivalently $U_G$) must be large enough so that at least one wavelength fits within the cross-stream size.

Equations (\ref{eq:conf}) or (\ref{eq:conf2}) can be applied to the data of figure \ref{fig:Marmottant}, for which the convective/confinement transition is observed. For the $\delta_d=1$ series of figure \ref{fig:Marmottant} left (symbol $\square$), the transition is predicted at $U_G=28$~m/s by equation (\ref{eq:conf2}) and observed at $U_G=27\pm5$~m/s in the full stability analysis. For the larger liquid velocity of figure \ref{fig:Marmottant} right (symbol $\square$), the transition for the same $\delta_d=1$ series is predicted at $U_G=49$~m/s by equation (\ref{eq:conf2}) and observed at $U_G=50\pm5$~m/s in the full stability analysis. The agreement is therefore good.

\begin{figure}
\centering
\includegraphics[width=0.45\textwidth]{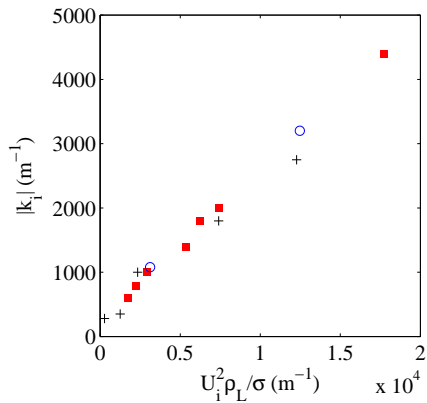}
\caption{The spatial growth rate at the pinch point with the surface tension branch is controlled by the interfacial velocity $U_i$ (red symbols of figures \ref{fig:Hg5Hl5} to \ref{fig:Marmottant}): ${\color{red}\blacksquare}$ $\delta_d=0.5$ ; $+$ $\delta_d=0.1$ ; the two cases with symbol ${\color{blue} \circ}$ correspond to fixed $U_G=39$~m/s, $U_L=0.45$~m/s, $\delta_d=0.5$ (hence fixed $U_i$) and to $\sigma=2\sigma_{water}$ and  $\sigma=\sigma_{water}/2$.}
\label{fig:ki_otto}
\end{figure}

We now turn to the question of the conditions for which the absolute instability triggered by surface tension is observed. This mechanism takes place when the most dangerous mode of the shear branch reaches the branch controlled by surface tension at lower $k_i$, see  the branch of figure \ref{fig:Hg5Hl5} right (${\color{red} +}$) and the numerous examples in  \citet{Otto}. This branch at low $k_i$ corresponds to capillary waves, which propagate upstream  (the group velocity $d\omega_r/dk_r$ is negative for this branch). In our problem the interface has a velocity $U_i$, both fluids are sheared, and the dispersion relation is therefore expected to be more complex than the classical dispersion relation of capillary waves (see for example the work of \citet{Young2014}, where a dispersion relation is derived for capillary waves in a similar context albeit with a different velocity profile and within an inviscid assumption). At any rate, the typical growth rate $k_{i\sigma}$ at which this branch is located is expected to depend on $\sigma/\rho_L$ and on the interfacial velocity. We plot on figure \ref{fig:ki_otto} the spatial growth rate at the pinch point $k_{i\sigma}$, corresponding to the red symbols of figures \ref{fig:Hg5Hl5} to \ref{fig:Marmottant}, as a function of $(\rho_L/\sigma)U_i^2$. We also include two artificial cases where $\sigma=2\sigma_{water}$ and  $\sigma=\sigma_{water}/2$, for fixed $U_G=39$~m/s, $U_L=0.45$~m/s, $\delta_d=0.5$, hence fixed $U_i$ (symbol ${\color{blue} \circ}$). Though a wide range of parameters are varied in these data, $k_{i\sigma}$ can be fairly well estimated by $k_{i\sigma}\sim (\rho_L/\sigma)U_i^2/4$. 

The surface tension absolute instability will therefore come into play when the spatial growth rate of the shear branch most dangerous mode $k_{i\;conv}$ reaches $k_{i\sigma}$, i.e. when:
$$\sqrt{\frac{\rho_G}{\rho_L}}\frac{U_G}{U_i}\left(\frac{\rho_G}{\sigma}\right)^{1/3}\left(\frac{U_G}{\delta_G}\right)^{2/3}>(\rho_L/\sigma)U_i^2/4$$

\begin{equation}
\frac{\rho_G^{5/6}\sigma^{2/3}}{\rho_L^{3/2}}\frac{U_G^{5/3}}{\delta_G^{2/3}}>U_i^3/4
\label{eq:otto_conv}
\end{equation}

Physically, the idea is that the surface tension resonance is favored by low interface velocities: capillary waves must be able to send information upstream. This is what is qualitatively observed in figures \ref{fig:Hg5Hl5} to \ref{fig:Marmottant}: when $U_G$ is increased at fixed $U_L$, the right hand side of equation \ref{eq:otto_conv} will increase faster than the left hand side ($\delta_G$ decreases at most as $U_G^{-1/2}$ for all geometries investigated), and the surface tension mechanism will necessarily give way to either the confinement or the convective regimes at large $U_G$. In addition, if at constant $U_G$ the velocity deficit is increased ($\delta_d$ decreased), $U_i$ is decreased, and the surface tension mechanism takes over. This is again what is observed in our data (see figures \ref{fig:Hg5Hl5} and \ref{fig:Marmottant}). 

We finally comment on the scaling laws for frequency in the three possible regimes. In the convective regime, we have shown previously that $f\sim \left(\rho_G/\sigma\right)^{1/3}\left(U_G/\delta_G\right)^{2/3}U_i$. If we assume $\delta_G\sim U_G^{-1/2}$, as is the case in most of our experimental configurations, this yields $f\sim U_GU_i$. This is consistent with the behavior observed on figure \ref{fig:Marmottant} right, where at fixed $U_L$ we have $f\sim U_G$ for low $U_G$. We do not discuss the scaling at large $U_G$ ($M\gg 1$), since for these conditions the convective regime necessarily gives way to the confinement regime.

In the surface tension controlled regime, the pinch point is always located close to the lowest $k_i$ reached by the shear branch. The frequency at the pinch point will therefore be close to the frequency of the most dangerous mode discussed above. The main difference with the convective regime resides in the nature of the instability: it is absolute, and the growth rate of the instability will therefore be controlled by non linear effects. In particular, wave velocity is expected to be close to $U_c =(\sqrt{\rho_G}U_G+\sqrt{\rho_L} U_L)/(\sqrt{\rho_G}+\sqrt{\rho_L})$  \citep{Dimotakis,Hoepffner},  which for air and water simplifies into $U_c \approx U_L+ \sqrt{\rho_G/\rho_L}U_G$. In the case of air and water, wave velocity is therefore much larger than the interfacial velocity, and the wavelength $\lambda\sim U_c/f$ much larger than in the convective regime.

In the confinement controlled absolute instability regime, the pinch point is located at $k_r\sim 1/L$. The corresponding frequency will then be a function of the phase velocity at low $k_r$.
As commented above, in the low $k_r$ limit the perturbation is controlled by a balance between the pressure perturbation on the gas side, proportional to $\rho_GU_G^2$, and  inertia on the liquid side. In inviscid approaches, this situation is classically associated with the  velocity $U_c$ introduced above. A velocity close to this one can be recovered in the present viscous context by writing the balance of normal stresses (equation \ref{norm_stress_cont}) at the interface for low wave numbers, balance between the pressure disturbance generated by vorticity on the less viscous (gas) side \citep{Hinch} and liquid inertia:
$$\mu_G\phi_G^{(''')}\sim-\phi_L'i\rho_L\left(kU_i-\omega\right)
+\phi_Li\rho_LkU_L'$$
$$\mu_G\rho_G\frac{k^2U_G}{\mu_G}\phi_G'\sim \rho_L\left(kU_L'+\frac{\omega-kU_i}{\delta_L}\right)\phi_L$$
$$\rho_G\frac{k^2U_G^2}{(\omega-kU_i)\delta_G}  \sim \rho_L\frac{\omega-kU_L}{\delta_L}$$
$$\frac{\omega}{k}\sim \sqrt{\frac{\rho_G}{\rho_L}\frac{\delta_L}{\delta_G}}U_G +U_L$$
This velocity is similar to the velocity $U_c$ introduced above. In the above estimate we have neglected the additional contribution due to the difference between $U_i$ and $U_L$ (this is approximately valid at moderate $U_G$ for the density and viscosity values of air and water at ambient conditions). The frequency of the instability can then be estimated by:
\begin{equation}
f\sim \frac{\sqrt{\frac{\rho_G }{\rho_L}\frac{\delta_L}{\delta_G}}U_G +U_L}{L}
\label{eq:conv}
\end{equation}
where $L$ is the cross-stream length controlling the confinement branch.
This expression predicts the correct trend and order of magnitude for the data of figures \ref{fig:Hg5Hl5} to \ref{fig:Marmottant} (blue symbols). Equation (\ref{eq:conv}) is also consistent with the decrease in frequency when the width of the nozzle is increased, an impact that was commented in the previous section: it decreases frequency by moving the confinement branch closer to the real axis. This effect is certainly the reason why the frequency measured by \citet{Marmottant} on their coaxial jet was twice as large as the frequency measured by \citet{Raynal_turb} on a planar mixing layer experiment for the same fluids and similar velocities but different geometries: the largest cross stream length scale in the coaxial jet experiment (4 mm jet radius) was approximately twice as small as the largest lengthscale in the mixing layer experiment (stream thickness 1 cm). The fact that the jet radius and not its diameter is the relevant length scale for the coaxial geometry is a consequence of the boundary condition of necessary zero velocity perturbation at the jet center.

In the previous normal stress balance, we have not included the liquid contribution $\mu_L\phi_L^{(''')}$ to the pressure disturbance at the interface. With this liquid contribution, and assuming $\delta_d=1$, the normal stress balance at the interface becomes:
$$\rho_G\frac{k^2U_G^2}{(\omega-kU_i)\delta_G}
+\rho_L\frac{k^2(U_i-U_L)^2}{(\omega-kU_i)\delta_L}\sim \rho_L\frac{\omega-kU_L}{\delta_L}$$
$$\frac{\omega^2}{k^2}\sim \frac{\rho_G}{\rho_L}\frac{\delta_L}{\delta_G}U_G^2 +\frac{\mu_G}{\mu_L}\frac{\delta_L^2}{\delta_G^2}U_G^2+U_L^2$$
For air and water, as long as $\delta_G\sim \delta_L$ the additional liquid contribution is negligible and the scaling of equation (\ref{eq:conv}) is expected to hold. However at large $U_G$, $\delta_G$ may eventually decrease to a point where the liquid contribution will eventually dominate. In this $\delta_G\ll\delta_L$ limit one will then recover the $f\sim U_G/\delta_G \sim U_G^{3/2}$ scaling law observed experimentally at large $U_G$.

For $\delta_d<1$, one can show that there is an additional term in the liquid contribution: this added vorticity is consistent with the increase in frequency when $\delta_d$ is decreased at constant $U_G$ and $U_L$ in figures \ref{fig:Hg5Hl20} to \ref{fig:Marmottant}.

\begin{figure}
\centering
\includegraphics[width=0.45\textwidth]{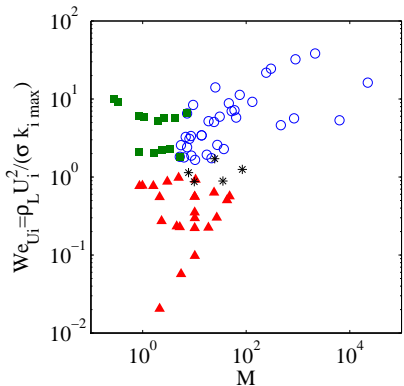}
\caption{Cartography of the instability regimes encountered in the present paper for the air/water coaxial jet in the $(M,We_{Ui})$ plane: ${\color{red} \blacktriangle}$ for the surface tension mechanism, ${\color{blue} \circ}$ for the confinement mechanism, and ${\color{green} \blacksquare}$ for the convective mechanism. Symbol $*$ corresponds to conditions where the confinement and surface tension instabilities have similar absolute growthrates.}
\label{fig:carto}
\end{figure}

The aim stated at the beginning of this paper was to clarify the boundaries between the three instability regimes, and draw a cartography for the case of the air/water coaxial jet. We now wish to confront the two conditions introduced previously, namely equations (\ref{eq:conf}) and (\ref{eq:otto_conv}), against all the stability analysis data presented in this paper. We first need to identify which dimensionless groupings have to be put on the axes of this cartography. Condition (\ref{eq:conf}) is expressed as a function of the dynamic pressure ratio $M$, we therefore retain this parameter. The idea behind equation (\ref{eq:otto_conv}) is that the Weber number $We_{Ui}$ built with the interfacial velocity and the most unstable wavelength $k_{r\;max}$ as a lengthscale must be smaller than one.
In order to test how these two criteria are consistent with the three regimes observed in our stability analysis, we therefore represent on figure \ref{fig:carto} all the cases presented in this paper (stability analysis data of figures \ref{fig:Hg5Hl5} to \ref{fig:ki_otto})   in the $(M,We_{Ui})$ plane. Each symbol corresponds to a given mechanism: ${\color{red} \blacktriangle}$ for the surface tension mechanism, ${\color{blue} \circ}$ for the confinement mechanism, and ${\color{green} \blacksquare}$ for the convective mechanism. We also represent with symbol $*$ the conditions where the confinement and surface tension instabilities have similar absolute growthrates. Figure \ref{fig:carto} indicates that the two dimensionless numbers identified in this paper are indeed relevant, since there is very little overlap between the cases corresponding to different mechanisms. The surface tension mechanisms appears to act for $We_{Ui}=\rho_LU_i^2/(\sigma k_{r\;max})<1$. The two other mechanisms only exist when $We_{Ui}=\rho_LU_i^2/(\sigma k_{r\;max})>1$: when $M<5$ the convective regime dominates, when $M>5$ the confinement absolute mechanism takes place.
  
\section{Conclusion}
We have shown via experiments and stability analysis that the shear instability between a slow  water jet and a coaxial fast air jet displays different regimes of instability depending on the destabilizing mechanisms: a convective instability  whose most dangerous mode results from a balance between shear in the gas stream and the cut-off of surface tension ; an absolute instability when interfacial velocity is low enough so that the shear branch can resonate with a capillary branch ; or an absolute instability controlled by confinement if the spatial growth rate of the shear instability is large enough to trigger a cross stream resonance. We have shown that these three distinct mechanisms compete with each other in a narrow range of experimental conditions in air/water experiments when the geometry and the liquid/gas velocities are varied: a change of regime can be detected by non monotonous variations of the frequency with $U_G$. We have introduced criteria to help predict which mechanism is expected to  dominate for given conditions. These criteria are given by equations (\ref{eq:conf}) and (\ref{eq:otto_conv}). The basic idea is that the confinement regime takes over the convective regime when the dynamic pressure ratio $M$ becomes large ; the condition for the occurrence of the surface tension absolute instability is more complex, and hence more difficult to summarize in an intuitive condition. 
It can be simplified in the limit when $We_{\gamma}>1$, when the most unstable mode of the shear branch is simply $k_{r\;max}=\delta_L^{-1}$: in this case the condition becomes $(\rho_L/\sigma)U_i^2\delta_L<4$. The fact that this Weber number built with the interfacial velocity and the liquid boundary layer thickness must be small can be understood as the interfacial speed $U_i$ being smaller than $\sqrt{\sigma/(\rho_L\delta_L)}$ which is analogous to a Taylor-Culick velocity. At any rate, the condition states that capillary waves must be able to send information upstream.

The question of the competition between both absolute instabilities has not been addressed. The cartography of figure \ref{fig:carto} suggests a simple picture: as soon as $We_{Ui}<1$ the surface tension instability dominates, and it seems that the confinement mechanism can only exist when the surface tension one is invalidated. However, a thorough discussion of this competition should include a discussion of the values of the absolute growth rates in each regime, and this is something that remains to be addressed.

Finally, we have proposed scaling laws for the most unstable frequency in each  regime: these are discussed at the end of the previous section, and are consistent with experimental observations.

In the present work we have limited ourselves to the study of axisymmetric perturbations close to the nozzle. However, and as mentioned in the first section, non-axisymmetric perturbations may eventually dominate the jet dynamics over short downstream distances. This then leads to large scale oscillations and subsequent break-up of the liquid jet (see figure \ref{fig:injector}c). This ``flapping instability'' of the liquid jet will be described in future work.

 We have shown that the magnitude of the velocity deficit in the base flow profile was crucial in controlling which mechanism is dominant, and the choice of the relevant $\delta_d$ coefficient is therefore the main source of ambiguity in the proposed stability analysis. For all the nozzles studied in this paper, the vorticity thicknesses $\delta_d\delta_L$ for which there is agreement between stability analysis and experimental data fall in the range 200-500 $\mu$m, which is the order of magnitude of the splitter plate thickness of the nozzles in the experiments, but it is difficult to conclude as to whether this is significant or just a coincidence. The main limitation of our analysis resides in the parallel flow assumption: in the experiment the velocity profile, deficit and radius actually change over distances of the order of the wavelength. A global stability analysis \citep{Huerre} could be a way to capture unambiguously the strong impact of the velocity profile on the developing flow. 
 
 A further issue, of particular relevance in applications, is the question of the impact of turbulence which is known to  affect stability \citep{ONaraigh,matas_turb} and therefore the scaling laws for frequency. Turbulence is expected to impact as well the competition between the instability regimes identified in the air-water configuration.\\

The LEGI laboratory is part of the LabEx Tec 21 (Investissements d'Avenir – grant agree\-ment n$^\circ$ ANR-11-LABX-0030).

\appendix
\section*{Appendix: Energy budget}
\label{annex:energy}

We give here the expressions for each of the terms in the energy budget of equation (\ref{energy_budget}). The expressions have the dimension of a local energy rate per unit axial length $x$.

\begin{align}
\notag
\frac{dE_{kin}}{dt}=2\pi\omega_i\left[ 
\rho_L\int_{0}^{R}\left(\phi_L'^2+ |k|^2\phi_L^2 \right)\frac{1}{r}dr+\rho_G\int_R^{L_G} \left(\phi_G'^2+ |k|^2\phi_G^2 \right)\frac{1}{r}dr\right]e^{2\omega_it-2k_ix}
\\
\notag
-2\pi k_i\left[\rho_L\int_{0}^{R}\left(\phi_L'^2+ |k|^2\phi_L^2 \right)\frac{U(r)}{r}dr+\rho_G\int_R^{L_G} \left(\phi_G'^2+ |k|^2\phi_G^2 \right)\frac{U(r)}{r}dr\right]e^{2\omega_it-2k_ix}
\end{align}

$$REY_L= \frac{i2\pi k\rho_L}{2}\left[
\int_{0}^{R}\left(\phi_L'^*\phi_L-\phi_L^*\phi_L'\right)
\frac{1}{r}\frac{dU}{dr}dr\right]e^{2\omega_it-2k_ix}
$$

$$REY_G= \frac{i2\pi k\rho_G}{2}\left[
\int_{R}^{L_G}\left(\phi_G'^*\phi_G-\phi_G^*\phi_G'\right)
\frac{1}{r}\frac{dU}{dr}dr\right]e^{2\omega_it-2k_ix}
$$

\begin{align}
\notag
DIS_L=-2\pi\mu_l\left[\int_{0}^R \left(
4|k|^2\left|\phi_L'\right|^2+\left|\phi_L''\right|^2+
\left(\phi_L''^*k^2\phi_L+k^{2*}\phi_L^*\phi_L''\right)+
\left|k\right|^4\left|\phi_L\right|^2 
\right.\right.
\notag\\
\notag
\left.\left.
+4\left|k\right|^2\frac{\left|\phi_L\right|^2}{r^2}-2\frac{\left|k\right|^2}{r}\left(\phi_L'^*\phi_L+\phi_L^*\phi_L'\right)
-\frac{1}{r}\left(\phi_L'^*k^2\phi_L+k^{2*}\phi_L^*\phi_L'\right)
\right.\right.
\notag\\
\notag
\left.\left.
+\frac{\left|\phi_L'\right|^2}{r^2}-\frac{1}{r}
\left(\phi_L''^*\phi_L'+\phi_L'^*\phi_L''\right)
\right)
\frac{1}{r}dr\right]e^{2\omega_it-2k_ix}
\end{align}

\begin{align}
\notag
DIS_G=-2\pi\mu_g\left[\int_{R}^{L_G} \left(
4|k|^2\left|\phi_G'\right|^2+\left|\phi_G''\right|^2+
\left(\phi_G''^*k^2\phi_G+k^{2*}\phi_G^*\phi_G''\right)+
|k|^4\left|\phi_G\right|^2 
\right.\right.
\notag\\
\notag
\left.\left.
+4|k|^2\frac{\left|\phi_G\right|^2}{r^2}
-2\frac{\left|k\right|^2}{r}\left(\phi_G'^*\phi_G+\phi_G^*\phi_G'\right)
-\frac{1}{r}\left(\phi_G'^*k^2\phi_G+k^{2*}\phi_G^*\phi_G'\right)
\right.\right.
\notag\\
\notag
\left.\left.
+\frac{\left|\phi_G'\right|^2}{r^2}-\frac{1}{r}
\left(\phi_G''^*\phi_G'+\phi_G'^*\phi_G''\right)
\right)
\frac{1}{r}dr\right]e^{2\omega_it-2k_ix}
\end{align}

\begin{align}
\notag
PDV_L=4\pi k_i\int_0^{R}\tilde{p}\tilde{u}^*rdr=4\pi k_i\int_0^{R}\left[\rho_L\frac{\phi_L'}{kr}(\omega-kU)+\rho_L\frac{\phi_L}{r}\frac{dU}{dr}
\right.
\notag\\
\notag
\left.+\frac{i\mu_L}{kr}\left(k^2\phi_L'+\frac{\phi_L''}{r}-\phi_L'''-\frac{\phi_L'}{r^2}
\right)\right]\phi_L'^*e^{2\omega_it-2k_ix}
\end{align}

\begin{align}
\notag
PDV_G=4\pi k_i\int_R^{L_G}\tilde{p}\tilde{u}^*rdr=4\pi k_i\int_R^{L_G}\left[\rho_G\frac{\phi_G'}{kr}(\omega-kU)+\rho_G\frac{\phi_G}{r}\frac{dU}{dr}
\right.
\notag\\
\notag
\left.+\frac{i\mu_G}{kr}\left(k^2\phi_G'+\frac{\phi_G''}{r}-\phi_G'''-\frac{\phi_G'}{r^2}
\right)\right]\phi_G'^*e^{2\omega_it-2k_ix}
\end{align}
where $\tilde{p}$ is the Fourier component of the pressure perturbation.

\begin{align}
\notag
TAN = \frac{\pi}{R}\left[\mu_g\left(\left(\phi_G'^*k^2\phi_G+k^{2*}\phi_G^*\phi_G'\right)
+\left(\phi_G''^*\phi_G'+\phi_G'^*\phi_G''\right)
-2\frac{\left|\phi_G'\right|^2}{R}\right)\right.
\\
\notag
\left.-\mu_l\left(\left(\phi_L'^*k^2\phi_L+k^{2*}\phi_L^*\phi_L'\right)
+\left(\phi_L''^*\phi_L'+\phi_L'^*\phi_L''\right)
-2\frac{\left|\phi_L'\right|^2}{R}\right)
\right]_{r=R}e^{2\omega_it-2k_ix}
\end{align} 

\begin{align}
\notag
NOR = k^2\left|\phi_G\right|^2\frac{i2\pi\sigma}{kU_i-\omega}\frac{1}{R^3}\left(1-k^2R^2\right)e^{2\omega_it-2k_ix}
\end{align}

\bibliographystyle{jfm}
\bibliography{prl_RMS_v3}

\end{document}